\documentclass[aps,prb,preprint,showpacs,superscriptaddress]{revtex4-1}

\usepackage{amsmath}
\usepackage{amssymb}
\usepackage{bm,color}
\usepackage{graphicx,epsfig,color}
\begin{document}

\title{Tailoring optical response of a hybrid comprising a quantum dimer emitter
strongly coupled to a metal nanoparticle}

\author{Bintoro S. Nugroho}
\affiliation{Zernike Institute for Advanced Materials, University of Groningen, Nijenborgh 4, 9747 AG Groningen, The Netherlands}
\affiliation{Fakultas Matematika dan Ilmu Pengetahuan Alam, Universitas Tanjungpura, Jl. Jendral A. Yani, 78124 Pontianak, Indonesia}

\author{Victor A. Malyshev}
\affiliation{Zernike Institute for Advanced Materials, University of Groningen, Nijenborgh 4, 9747 AG Groningen, The Netherlands}

\author{Jasper Knoester}
\affiliation{Zernike Institute for Advanced Materials, University of Groningen, Nijenborgh 4, 9747 AG Groningen, The Netherlands}

\date{\today}

\begin{abstract}
We study theoretically the optical response of a nanohybrid comprising a symmetric quantum dimer emitter coupled to a metal nanoparticle (MNP). The interactions between the exitonic transitions in the dimer and the plasmons in the MNP lead to interesting effects in the composite's input-output characteristics for the light intensity and the absorption spectrum, which we study in the linear and nonlinear regimes. We find that the exciton-plasmon hybridization leads to optical bistability and hysteresis for the one-exciton transition and enhancement of excitation for the two-exciton transition. The latter leads to a significant decrease of the field strength needed to saturate the system. In the linear regime, the absortion spectrum has a dispersive (Fano-like) line shape.
The spectral position and shape of this spectrum depend on the detuning of the dimer's one-exciton resonance relative to the plasmon resonance and the ratio of the exciton-plasmon coupling constant and the exciton dephasing rate.
Upon increasing the applied field intensity to the nonlinear regime, the Fano-like singularities in the absorption spectra are smeared and they disappear due to the saturation of the dimer, which leads to the MNP dominating the spectrum. The above effects, for which we provide physical explanations, allow one to tailor the Fano-like shape of the absorption spectrum, by changing either the detuning or the input power.
\end{abstract}


\maketitle

\section{Introduction}
\label{Introduction}

Over the past decade, heterogeneous nanosystems, consisting of quantum emitters (quantum dots, molecules, molecular aggregates, etc.) in close proximity to metal surfaces~\cite{SalomonPRL2012,SukharevACSnano2012} or metal nanoparticles (MNPs), ~\cite{ZhangPRL2006,VisteACSNano2010,ChenPRL2013,ZenginSciRep2013,PaspalakisJAP2014} have received a great deal of attention. The intriguing features of these systems arise from the hybridization of different types of optical excitations, in particular, excitons (in the quantum emitter) and plasmons (in the metal). The exciton-plasmon coupling can drastically modify the optical response of hybrids, leading to interesting physical phenomena, such as optical bistability,~\cite{ArtusoNanoLett2008,MalyshevPRB2011,ArtusoPRB2011,LiOptExpress2012,NugrohoJCP2013}
exciton-induced transparency~\cite{ArtusoPRB2010} enhancement of Rabi oscillations,~\cite{SadeghiNanotechnology2009}
suppression of quantum coherence via infrared driven coherent exciton-plasmon coupling,~\cite{SadeghiAPL2014,SadeghiNanotech2015}
florescence,~\cite{GuzatovJPCC2012,KulakovichNanoLett2002} F{\"o}rster energy transfer,~\cite{KyasJCP2011,MarocicoPRA2011,ZhangACSNano2012,ZhangACSNano2014} photoluminescence quenching,~\cite{PonsNanoLett2007} 
Fano-like absorption,~\cite{WiederrechtNanoLett2004,UwadaJPCC2007,KelleyNanoLett2007,ArtusoNanoLett2008,ZhangPRB2011,DeinegaJCP2014}
and other exciting effects.~\cite{FofangNanoLett2011, AberraPRL2012,ArtusoPRB2013,SlowikPRB2013,MarunPlasmonic2014,CarrenoJAP20014} These phenomena may have a strong impact on the development of active nanophotonic devices and metamaterials (e.g., optical switches, single photon sources, biosensors, etc.).

In this paper, we theoretically investigate the optical response of a nanocomposite consisting of a symmetric dimer of two two-level molecules coupled to an MNP. This work is an extension of works performed earlier by us~\cite{NugrohoJCP2013,MalyshevPRB2011} and other authors,~\cite{ArtusoNanoLett2008,LiOptExpress2012} on the optical response of single two-level molecules coupled to an MNP. These earlier studies revealed the interesting possibility of bistable optical response for such nanocomposites, arising from the self-action of the molecular excitation on itself through the reflection of the electric field generated by it on the MNP. This is a direct consequence of exciton-plasmon coupling. 

The dimer is a first step to considering molecular aggregates coupled to MNPs, a topic that recently has attracted experimental interest.~\cite{vujačićJPCC2012,DelacyOE2013,ZenginSciRep2013}
In contrast to a single two-level molecule, a dimer represents an optical ladder system consisting of a ground state, two one-exciton states, and one two-exciton state.  Direct (one-photon) transitions between the ground state and the two-exciton state are not allowed. By their richer excited-state structure, dimers coupled to MNPs may be expected to reveal new effects in their optical response, as compared to single molecules coupled to MNPs. Examples inspired by effects observed for ladder type (three-level) quantum dots are two-photon Rabi oscillations~\cite{StuflerPRB2006,MachnikowskiPSSC2008} and suppression of quantum decoherence.~\cite{SadeghiAPL2014,SadeghiNanotech2015} Another effect that in principle could occur is multi-stable behavior.~\cite{AsadpourOC2014}

We focus on two types of aspects of the optical response of the hybrid. The first one is the intensity dependent response at a few selected driving frequencies.  Quantities of interest are the occupation probabilities of various exciton levels in the system and the expectation value of the dipole of the dimer. We will show that the one-exciton manifold gives rise to bistable response and hysteresis in these quantities, while the two-exciton manifold does not. No multistability is found. We also show that two-photon absorption to the two-exciton state is strongly enhanced by the exciton-plasmon coupling, an effect that also has been observed experimentally~\cite{SivapalanLang2012} and which is of interest to two-photon microscopy. This enhancement also leads to a low saturation intensity of the dimer as compared to the case where it is not coupled to a MNP.

The second aspect we investigate is the absorption spectrum of the hybrid system. Most of the results obtained in this part do not depend on the presence of a two-exciton level. Yet, our work extends beyond previous results obtained for single molecules coupled to an MNP, because a much more detailed study is carried out than before, considering more different conditions.  In doing so, we show that the shape of the absorption spectrum in general strongly depends on the conditions, in particular, on the detuning between the one-exciton resonance and the plasmon resonance and on the ratio between the effective self-interaction parameter of the dimer and the dephasing rate of its optical transitions. In the weak-field limit, where only the one-exciton transitions play a role, the shape generally is dispersive and for a range of exciton-plasmon detunings is Fano-like; close to zero detuning, this shape disappears and a deep absorption dip may occur in the broad plasmon absorption line. In the strong-field limit, the dimer easily saturates and the Fano-like shape does not occur. These results show the sensitivity of the exciton-plasmon hybridization to precise conditions and accentuate the possibility to tailor the absorption line shape through appropriate system preparation.

This paper is organized as follows. In the next section, we present the model and the theoretical background to treat the optical response of a dimer-MNP hybrid. In Sec.~\ref{FieldDep}, we discuss the field dependent optical response both of an isolated dimer and a dimer-MNP composite (bistability and hysteresis). In Sec.~\ref{Spect}, the Fano-like absorption spectrum of the hybrid is discussed, both in the limit of weak and high applied field intensity. We summarise in Sec.~\ref{summary}.

\section{Model and theoretical background}
\label{setup}

\begin{figure}[ht]
\begin{center}
\includegraphics[width=0.8\columnwidth]{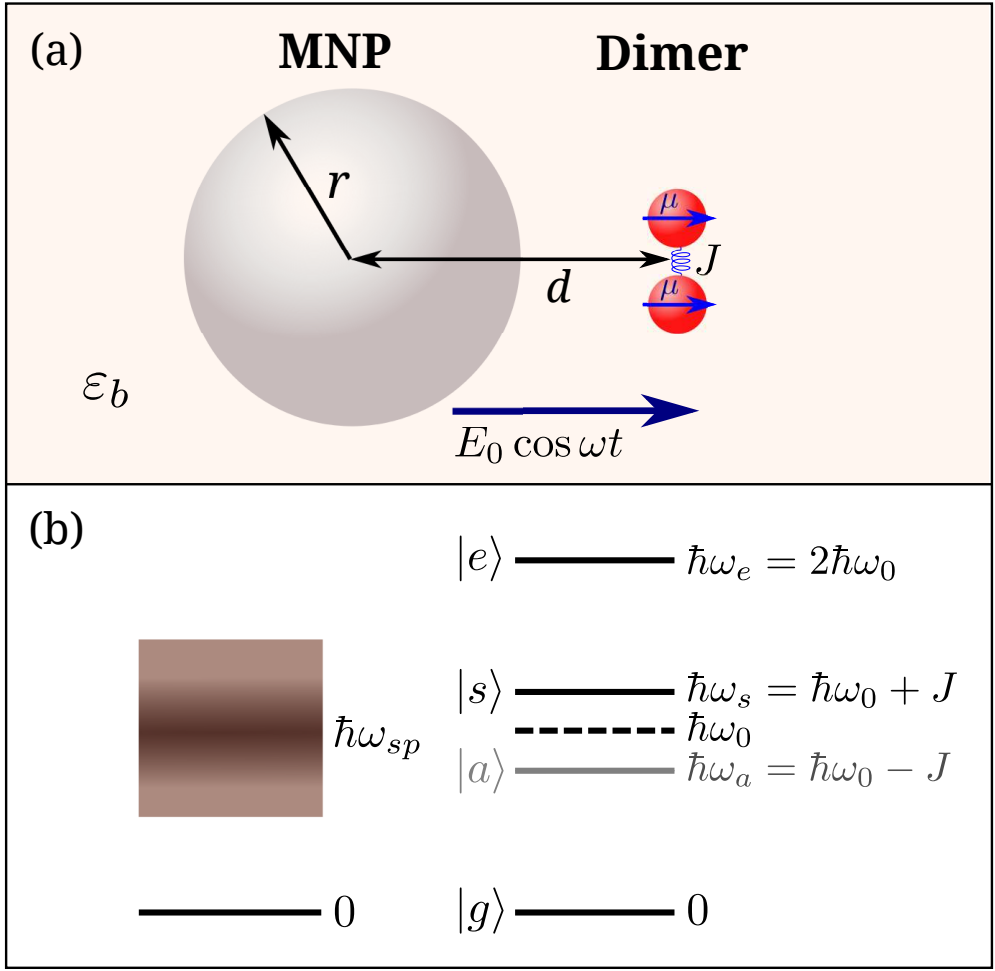}
\end{center}
\caption{(a)~Schematics of a hybrid system of a molecular dimer at a distance $d = |{\bf d}|$ from a MNP subject to an applied field $E_0\cos\omega t$ polarized along ${\bf d}$. The dimer consists of two identical quantum emitters with equal transition energies $\hbar \omega_0$ and transition dipole moments $\boldsymbol{\mu}$ and coupling strength $J$. The system is embedded in a homogeneous dielectric host with permittivity $\varepsilon_b$.  (b)~Diagram of the continuum of energy levels of the MNP (left) and the collective (excitonic) levels of the dimer (right).}
\label{Fig1}
\end{figure}

We consider the optical response of a hybrid consisting of a dimer of quantum emitters (two level molecules) coupled to a nearby spherical MNP (Ag) embedded in a dielectric (dispersionless) background with permittivity $\varepsilon_b$. The entire system is subjected to an external field of amplitude $\boldsymbol{E}_0$ and frequency $\omega$, $\bm{\mathcal{E}}_0(t) = \boldsymbol{E}_0(t) \cos (\omega t)$, oriented along the vector ${\bf d}$ between the centers of the MNP and the dimer. Figure~\ref{Fig1}(a) schematically shows the configuration. The radius $r$ of the MNP and the MNP-dimer spacing, $d$, are assumed to be small compared to the optical wavelength, thus allowing the use of the quasi-static approximation and the point dipole approximation for both particles~\cite{bohren2008,maier2007}. Also, the effects of retardation are negligible: the interaction between the the MNP and the dimer is dominated by the near-field dipole dipole coupling. For the radii of the MNP we consider ($r > 10~\mathrm{nm}$), effects of size quantization of the plasmon spectrum is negligible~\cite{SchollNat2012} (see Appendix A). Finally, we assume that there is no charge exchange between the dimer and the MNP; for $d-r < 2$ nm this might occur due to orbital overlap.~\cite{RidolfoPRL2010}

Like most of the previous work on these types of nanohybrids, we adopt the semiclassical approximation, i.e., we treat the optical response of the molecules quantum mechanically, while the response of the plasmons in the MNP is treated classically, using its polarizability. A full quantum mechanical treatment of the many-electron system that comprises the MNP is beyond the scope of this paper. Quantum models treating the MNP essentially with a few quantum degrees of freedom recently have been studied~\cite{ZhangPRB2011,SlowikPRB2013}; while such models are interesting, they do not describe effects such as bistable optical response. Thus, we describe the MNP by means of its frequency dependent polarizability:
\begin{equation}
  \alpha(\omega)= 4\pi r^3 \frac{\varepsilon_m(\omega)-\varepsilon_b}
                  {\varepsilon_m(\omega) + 2\varepsilon_b}~,
\label{alpha}
\end{equation}
where $\varepsilon_m(\omega)$ stands for the dielectric function of the metal. The frequency at which the real part of $\alpha(\omega)$ is minimal determines the plasmon resonance $\omega_{sp}$ (Fr\"{o}hlich condition). Thus, the electronic states of the MNP can be described by a ground state and a continuum of excited states with a well defined plasmon resonance $\omega_{sp}$  [See Fig.~\ref{Fig1}(b), left panel].
For the size of our interest corrections to $\alpha(\omega)$ due to the depolarization shift and radiative damping~\cite{meierOptlett1983} are negligible. Likewise, chemical interface damping which important for radii below 10 nm can safely  be neglected.~\cite{maier2007}

We model the dimer as two identical two-level quantum emitters, labeled $j=1,2$, with ground and excited states indicated as $|jg\rangle$ and $|je\rangle$, respectively; the emitters both have a resonance frequency $\omega_0$ and equally oriented transition dipole moments $\boldsymbol{\mu}$ which are, for definiteness, chosen perpendicular to the dimer axis (see Fig.~\ref{Fig1}); they have no static dipole moment. These two emitters interact resonantly with each other with coupling strength $J > 0$, resulting in four collective (excitonic) eigenstates: the ground state $|g\rangle = |1g,2g\rangle$, an antisymmetric and a symmetric one-exciton state, defined through $|a\rangle=1/(\sqrt{2})(|1g,2e\rangle - |1e,2g\rangle)$ and $|s\rangle=1/(\sqrt{2})(|1g,2e\rangle + |1e,2g\rangle)$, respectively, and the doubly excited state $|e\rangle=|1e,2e\rangle$.  The energies of these states are $0$, $\hbar\omega_{a} = \hbar\omega_0 - J$, $\hbar\omega_{s} = \hbar\omega_0 + J$, and $\hbar\omega_e = 2\hbar\omega_0$, respectively [see Fig.~\ref{Fig1}(b), right panel]. For the chosen geometry of the dimer, indeed the dipople-dipole coupling $J$ is positive, so that the symmetric state is the higher state of the one-exciton doublet. The antisymmetric state $|a\rangle$ is dark (zero transition dipole moment to the ground state) and does not contribute to the system's optical response, whereas the symmetric one has transition dipole moment $\boldsymbol{D} = {\sqrt 2}\boldsymbol{\mu}$. The transition between the doubly excited state $|e \rangle$ and the symmetric one also has the transition dipole moment $\boldsymbol{D}$. Transitions between the dimer's ground state and doubly excited state are one-photon forbiddden and can only be achieved by absorption or emission of two photons. Thus, optically, this symmetric dimer represents a three-level ladder-like quantum system.

We describe the optical dynamics of the dimer within the framework of the density matrix formalism. As all dipoles and electric fields are parallel to each other, they can be considered as scalars. Then, the set of equations for the density matrix elements $\rho_{gg}$, $\rho_{ss}$, $\rho_{ee}$, $\rho_{sg} = \rho_{gs}^*, \rho_{es} = \rho_{se}^*$, and $\rho_{eg} = \rho_{ge}^*$ reads (see also Ref.~\onlinecite{MalyshevPRA1998})
\begin{subequations}
\label{allrho}
\begin{align}
\begin{split}
\dot{\rho}_{gg} &= 2\gamma \rho_{ss} + i \frac{D{\cal{E}}_{D}}{\hbar} \left( \rho_{sg} - \rho_{sg}^* \right)~,
\end{split}
\label{rhogg}\\
\begin{split}
\dot{\rho}_{ss} &=  2\gamma Z_{es} + i \frac{D{\cal{E}}_{D}}{\hbar}\left( \rho_{es} - \rho_{es}^*
                 + \rho_{sg}^* - \rho_{sg} \right)~,
\end{split}
\label{rhoss}\\
\begin{split}
\dot{\rho}_{ee} &= -2\gamma \rho_{ee} + i \frac{D{\cal{E}}_{D}}{\hbar}\left( \rho_{es}^* -\rho_{es} \right)~,
\end{split}
\label{rohee}\\
\begin{split}
\dot{\rho}_{sg}    &= - \left( i\omega_{sg} +\Gamma \right) \rho_{sg} + 2\gamma \rho_{es}
+ i\frac{D{\cal{E}}_{D}}{\hbar} \left( \rho_{gg} - \rho_{ss} + \rho_{eg} \right)~,
\end{split}
\label{rhosg}\\
\begin{split}
\dot{\rho}_{es}    &= - \left( i\omega_{es} + 2\Gamma \right) \rho_{es} - i\frac{D{\cal{E}}_{D}}{\hbar} \left( Z_{es} + \rho_{eg} \right)~,
\end{split}
\label{rhoes}\\
\begin{split}
\dot{\rho}_{eg}    &= - \left( i\omega_{eg} +\Gamma\right) \rho_{eg} + i\frac{D{\cal{E}}_{D}}{\hbar} ( \rho_{sg} - \rho_{es} )~.
\end{split}
\label{rhoeg}
\end{align}
\end{subequations}
Here, $\gamma$ is the population radiative rate of a single emitter. Rigorously
speaking, $\gamma$ should be renormalized (enhanced) to $(1+F)\gamma$, where $F$ is the Purcell enhancement factor due to the presence of the closely spaced MNP. In Ref.~\onlinecite{CarminatiOptCommun2006}, it has been shown that $F$ is proportional to the real part of the polarizability $\alpha(\omega)$. Here, the actual frequency range we are interested in is around the MNP plasmon resonance, where the real part of $\alpha(\omega)$  tends to zero. Therefore, the Purcell effect is not important in our case, allowing us to take for the radiative rate its free-space value.
$\Gamma = \gamma + \Gamma^{\prime}$, with $\Gamma^{\prime}$ denoting the pure dephasing rate of the coherences (off-diagonal density matrix elements); $\omega_{sg}$, $\omega_{es}$, and $\omega_{eg}$ are the frequencies of the corresponding transitions: $|s\rangle \rightarrow |g\rangle$, $|e\rangle \rightarrow |s\rangle$, and $|e \rangle \rightarrow |g\rangle$; $Z_{sg} = \rho_{ss} - \rho_{gg}$ and $Z_{es} = \rho_{ee} - \rho_{ss}$ are the population differences between the states indicated in the subscripts. Furthermore, $\cal{E}_{D}$ denotes the electric field acting on the dimer; this field is the sum of the external field ${\cal{E}}_{0} = E_{0} \cos (\omega t)$ and the near field produced by the MNP at the position of the dimer:

\begin{equation}
\label{calE}
{\cal E}_{D} = {\cal E}_0 + \frac{{\cal P}_{MNP}}{2\pi \varepsilon_0 \varepsilon_b d^3}~,
\end{equation}
where $\varepsilon_0$ is the vacuum dielectric constant and ${\cal P}_{MNP}$ is the MNP's dipole moment induced by the external field ${\cal E}_0$ and the field produced by the dimer at the position of the MNP:
\begin{equation}
\label{calPMNP}
{\cal P}_{MNP} = \frac{1}{2} \varepsilon_0 \varepsilon_b \alpha (\omega)\left[ E_0 e^{-i \omega t} +
\frac{D}{\pi \varepsilon_0 \varepsilon_b d^3} ( \rho_{sg} + \rho_{es}) \right] + \mathrm{c.c.}~.
\end{equation}
Here the second term in square brackets comes from the field produced by the dimer dipole at the position of the MNP:
\begin{equation}
\label{calPD}
{\cal P}_{D} = D (\rho_{sg} + \rho_{es} + \mathrm{c. c.})
\end{equation}
From Eqs.~(\ref{calE}) - (\ref{calPD}), we obtain the following expression for the field acting on the dimer:
\begin{equation}
\label{calED}
{\cal E}_{D} = \frac{1}{2} \left[ 1 + \frac{\alpha(\omega)}{2\pi d^3} \right] E_0 e^{-i\omega t}
+ \frac{\alpha(\omega)D}{4\pi^2\varepsilon_0 \varepsilon_b d^6} (\rho_{sg} +\rho_{es}) + \mathrm{c. c.}
\end{equation}
The term in square brackets in Eq.~(\ref{calED}) describes the renormalization of the external field amplitude $E_0$ due to the presence of the MNP, while the second one is the self-action of the dimer via the MNP. As will be shown below, the latter may drastically modify the optical response of the composite compared to its separate components.

For the case of $\omega \approx \omega_0$ and the exciton splitting $2J/\hbar \ll \omega$, $\omega_0$, the rotating wave approximation can safely be used to simplify the treatment. To this end, we substitute in Eqs.~(\ref{rhogg}) -~(\ref{rhoeg}) the off-diagonal density matrix elements $\rho_{sg}, \rho_{es}$, $\rho_{eg}$, and the field ${\cal E}_D$ in the form
\begin{subequations}
\label{all_rho}
\begin{align}
\rho_{sg} &= R_{sg} e^{-i\omega t}~, \quad \rho_{es} = R_{es} e^{-i\omega t}~, \quad \rho_{eg} = R_{eg} e^{-2i\omega t}~,
\label{rho}\\
{\cal E}_D &= \frac{1}{2} E_D e^{-i\omega t} + \mathrm{c. c.}~,
\\
\intertext{with}
E_{D} &= \left[ 1 + \frac{\alpha(\omega)}{2\pi d^3} \right] E_0  + \frac{\alpha(\omega)D}{2\pi^2\varepsilon_0 \varepsilon_b d^6} (R_{sg} + R_{es})~,
\label{ED}
\end{align}
\end{subequations}
where the amplitudes $R_{sg}$,  $R_{es}$, $R_{eg}$, and $E_D$ are complex functions slowly varying on the time scale of $2\pi/\omega$.
After neglecting all terms oscillating with double frequency $2\omega$, the equations for the slow variables read:
\begin{subequations}
\label{all_dR1}
\begin{align}
\begin{split}
\dot{\rho}_{gg} &= 2\gamma \rho_{ss} + i \left( \Omega_D^* R_{sg} - \Omega_D R_{sg}^* \right)~,
\end{split}
\label{dRgg1}\\
\begin{split}
\dot{\rho}_{ss} &=  2\gamma Z_{es} + i \left( \Omega_D^* R_{es} - \Omega_D R_{es}^*
                 + \Omega_D R_{sg}^* - \Omega_D^* R_{sg} \right)~,
\end{split}
\label{dRss1}\\
\begin{split}
\dot{\rho}_{ee} &= -2\gamma \rho_{ee} + i \left( \Omega_D R_{es}^* -\Omega_D^* R_{es} \right)~,
\end{split}
\label{dRee1}\\
\begin{split}
\dot{R}_{sg}    &= - \left( \Gamma - i\Delta_{sg} \right) R_{sg} + 2\gamma R_{es} + i \left( \Omega_D^* R_{eg} - \Omega_D Z_{sg}\right)~,
\end{split}
\label{dRsg1}\\
\begin{split}
\dot{R}_{es}    &= - \left( 2\Gamma -i\Delta_{es} \right)R_{es} - i \left( \Omega_D Z_{es} + \Omega_D^* R_{eg} \right)~,
\end{split}
\label{dRes1}\\
\begin{split}
\dot{R}_{eg}    &= - \left( \Gamma - i\Delta_{eg} \right) R_{eg} + i\Omega_D ( R_{sg} - R_{es} )~.
\end{split}
\label{dReg1}
\end{align}
\end{subequations}
Here, $\Delta_{sg}=\omega-\omega_{sg}$, $\Delta_{es}=\omega-\omega_{es}$, and $\Delta_{eg} = 2\omega - \omega_{eg} = 2(\omega - \omega_0)$ are the detunings out of resonance of the one-photon $|s\rangle \leftrightarrow |g\rangle$ and $|e\rangle \leftrightarrow |s\rangle$, and two-photon $|e\rangle \leftrightarrow |g\rangle$ transitions, respectively. Furthermore, $\Omega_D = D E_D/(2\hbar)$
is given by
\begin{equation}
\label{OmegaD}
\Omega_D = {\widetilde{\Omega}}_0 + G (R_{sg} + R_{es})
\end{equation}
with
\begin{subequations}
\label{TildeOmega0G}
\begin{align}
\begin{split}
{\widetilde \Omega}_0 &= \left[ 1 + \frac{\alpha(\omega)}{2\pi d^3} \right] \Omega_0~,
\end{split}
\label{TildeOmega}\\
\begin{split}
G &= \frac{D^2 \alpha(\omega)}{4\pi^2 \hbar \varepsilon_0 \varepsilon_b d^6}~,
\end{split}
\label{G}
\end{align}
\end{subequations}
where $\Omega_0 = D E_0/(2\hbar) = \mu E_0/(\sqrt{2}\hbar)$ is the Rabi frequency of the external field. The complex-valued constant $G  =G_R + i G_I$ is the feed-back parameter arising from the dimer-dimer self-action via the MNP. It contains all details of the dimer-MNP interaction, material properties, and geometry of constituents.

It is useful to realize that $\hbar G = D^2\alpha(\omega)/(4\pi^2\epsilon_0\epsilon_b d^6)$ looks like the inductive dipole-dipole interaction between the dimer and the MNP, even if it scales with the dimer-MNP distance $d$ as $d^{-6}$; we use here the term "inductive" because an MNP, as a classical object, has neither a static nor a transition dipole moment. Up to coefficients of order unity and signs, the field produced by the dimer's transition dipole $D$ at the position of the MNP is $D/d^3$. This field induces a dipole $D\alpha/d^3$ in the MNP, which in turn has an interaction $D^2\alpha/d^6$ with the dimer's transition dipole, confirming our interpretation of $G$.

Substituting Eq.~(\ref{OmegaD}) and Eqs.~(\ref{TildeOmega}) -~(\ref{G}) into Eqs.(\ref{dRgg1}) -~(\ref{dReg1}), we obtain the equations of motion for the density matrix elements of the hybrid in the form

%%while
\begin{subequations}
\label{all_dR2}
\begin{align}
\begin{split}
\dot{\rho}_{gg} &= 2\gamma \rho_{ss} + i(\widetilde{\Omega}_0^* R_{sg} - \widetilde{\Omega}_0 R_{sg}^*) \\
          &\quad + i(G^*R_{es}^* R_{sg} - GR_{es} R_{sg}^*)+ 2G_I R_{sg}R_{sg}^*~,
\end{split}
\label{dRgg2}\\
\begin{split}
\dot{\rho}_{ss} &= 2\gamma Z_{es} + i\left[\widetilde{\Omega}_0^*(R_{es} -R_{sg}) + \widetilde{\Omega}_0(R_{sg}^*- R_{es}^*) \right] \\
          &\quad + 2iG_R(R_{sg}^*R_{es}-R_{sg}R_{es}^*) + 2G_I(R_{es}R_{es}^* - R_{sg}R_{sg}^*)~,
\end{split}
\label{dRss2}\\
\begin{split}
\dot{\rho}_{ee} &= -2\gamma\rho_{ee} - 2G_I R_{es}R_{es}^* + i(\widetilde{\Omega}_0 R_{es}^* - \widetilde{\Omega}_0^* R_{es})\\
          &\quad + i(GR_{sg}R_{es}^* - G^*R_{sg}^*R_{es}) ~,
\end{split}
\label{dRee2}\\
\begin{split}
\dot{R}_{sg}  &= -\left[ \Gamma - G_IZ_{sg} + i(G_RZ_{sg}-\Delta_{sg})  \right]R_{sg} + 2\gamma R_{es} +
i(\widetilde{\Omega}_0^*R_{eg} - \widetilde{\Omega}_0 Z_{sg} )\\
          &\quad + i\left[G^*(R_{es}^* + R_{sg}^*)R_{eg} - GZ_{sg}R_{es}\right]~,
\end{split}
\label{dRsg2}\\
\begin{split}
\dot{R}_{es}    &= - \left[ 2\Gamma - G_IZ_{es} + i(G_RZ_{es}-\Delta_{es})  \right]R_{es} - i(\widetilde{\Omega}_0Z_{es} + \widetilde{\Omega}_0^*R_{eg})\\
          &\quad - i\left[GZ_{es}R_{sg} + G^*(R_{es}^* + R_{sg}^*)R_{eg}\right]~,
\end{split}
\label{dRes2}\\
\begin{split}
\dot{R}_{eg}     &= - (\Gamma-i\Delta_{eg})R_{eg} + i\widetilde{\Omega}_0(R_{sg}-R_{es}) + iG(R_{sg}R_{sg}-R_{es}R_{es})~,
\end{split}
\label{dReg2}
\end{align}
\end{subequations}

As follows from Eqs.~(\ref{dRgg2}) -~(\ref{dReg2}), the dimer's self-action, governed by the feedback parameter $G$, gives rise to additional nonlinearities as compared to an isolated dimer. Two of these that should be mentioned in particular, are (i) - renormalization of the dimer's transition frequencies, $\omega_{sg} \rightarrow \omega_{sg} + G_R Z_{sg}$ and  $\omega_{es} \rightarrow \omega_{es} + G_R Z_{es}$, and (ii) - renormalization of the damping rates of the off-diagonal density matrix elements, $\Gamma \rightarrow \Gamma - G_I Z_{sg}$ and $2\Gamma \rightarrow 2\Gamma -G_I Z_{es}$, both depending on the corresponding population differences. As will be shown below, these two play an important role in understanding the hybrid's optical response. The nonlinearities introduced by these renormalizations are similar to those found for a two-level system close to a MNP,~\cite{ZhangPRL2006,MalyshevPRB2011,NugrohoJCP2013} except that more levels are involved here.

\section{Response at a fixed frequency}
\label{FieldDep}

\begin{figure}[h!]
\centering
\includegraphics[width=0.80\linewidth]{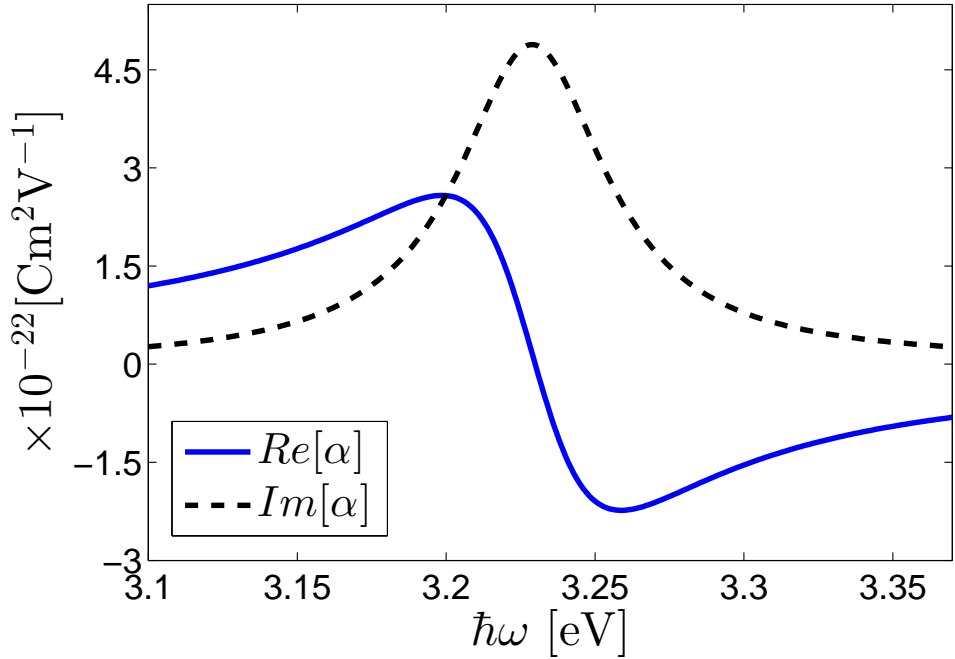}
\caption{Frequency dependence of the real ($\mathrm{Re[\alpha]}$) and imaginary ($\mathrm{Im[\alpha]}$) parts of the silver MNP polarizability $\alpha$, calculated according to Eq.~(\ref{alpha}) with the set of parameters given in the text.}
\label{polarizability}
\end{figure}

In this section, we investigate the optical response of the nanohybrid at a fixed frequency and are interested in the dependence on the magnitude of the driving field (i.e., the Rabi frequency $\Omega_0$). To this end, we numerically solve Eqs.~(\ref{dRgg2})~-~(\ref{dReg2}). In our calculations, we consider a spherical silver MNP of size $r = 11~\mathrm{nm}$ and use the dielectric function $\varepsilon_m(\omega)$ for Ag as obtained from experimental data.~\cite{JohnsonPRB1972}  For the host's dielectric constant we assume $\varepsilon_b=1.80$. Using these data and applying Eq.~(\ref{alpha}), the peak of the plasmon resonance is found around $\hbar\omega_{sp}=3.23~\mathrm{eV}$. Figure~\ref{polarizability} illustrates the full frequency dependence of the polarizability $\alpha(\omega)$ for this set parameters. For the two-level systems that make up the dimer, we assume a resonance frequency $\hbar\omega_0 = 3.13~\mathrm{eV}$, a transition dipole moment $\mu = 0.5~e \cdot\mathrm{nm}$ (with $e$ the magnitude of the electron charge), population and coherence relaxation rates of $\gamma = 1.43~\mathrm{ns}^{-1}$ ($\hbar\gamma = 0.94 \times 10^{-3}~\mathrm{meV}$) and $\Gamma' =
1.54~\mathrm{ns}^{-1}$ ($\hbar\Gamma^{\prime} = 10^{-3}~\mathrm{meV}$), respectively, and a mutual coupling of $J=0.1~\mathrm{eV}$.

We performed calculations of the dimer populations $\rho_{gg}, \rho_{ss}$, and $\rho_{ee}$ as well as the absolute value of the normalized amplitude of the dimer's dipole moment $|R_{es} + R_{sg}|$, sweeping adiabatically the magnitude of the applied field up and down. For comparison, we did this for both an isolated dimer and a dimer-MNP hybrid. The applied field frequency $\omega$ was set in resonance with the bright dimer transition $|g\rangle \rightarrow |s\rangle$, i.e., $\hbar\omega = \hbar\omega_{sg} = \hbar\omega_{0}+J = 3.23~\mathrm{eV}$. Note that for the chosen parameters $\omega_{sg}$ coincides with the plasmon resonance $\omega_{sp}$. The results are presented in Fig.~\ref{DiagOffDiagVsIntens_Isolated} (isolated dimer) and Fig.~\ref{DiagOffDiagVsIntens_Coupled} (hybrid) and will be discussed below.

\subsection{Isolated dimer}
\label{Isolated dimer}

\begin{figure}[ht]
\begin{center}
\includegraphics[width=0.80\columnwidth]{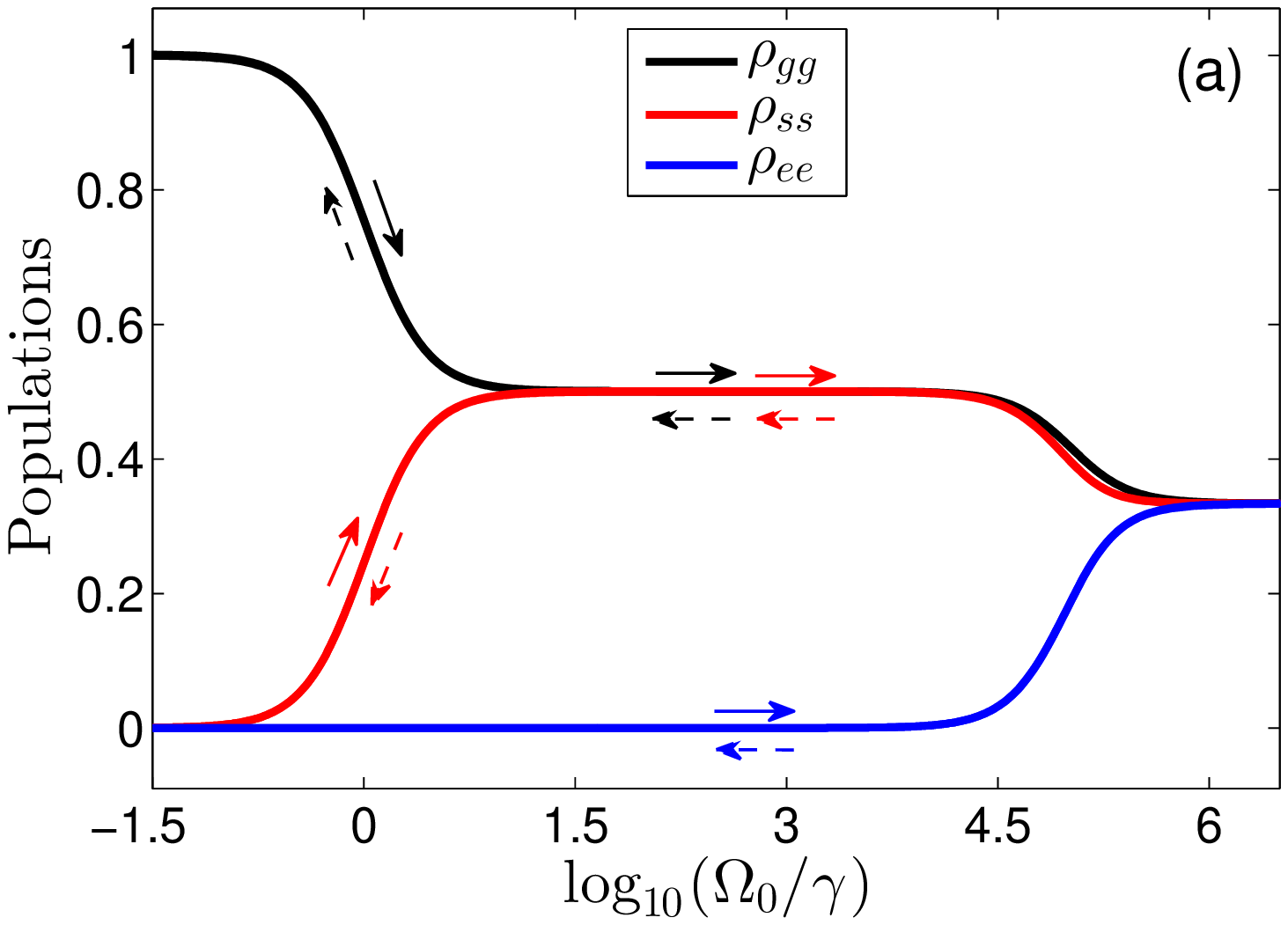}
\includegraphics[width=0.80\linewidth]{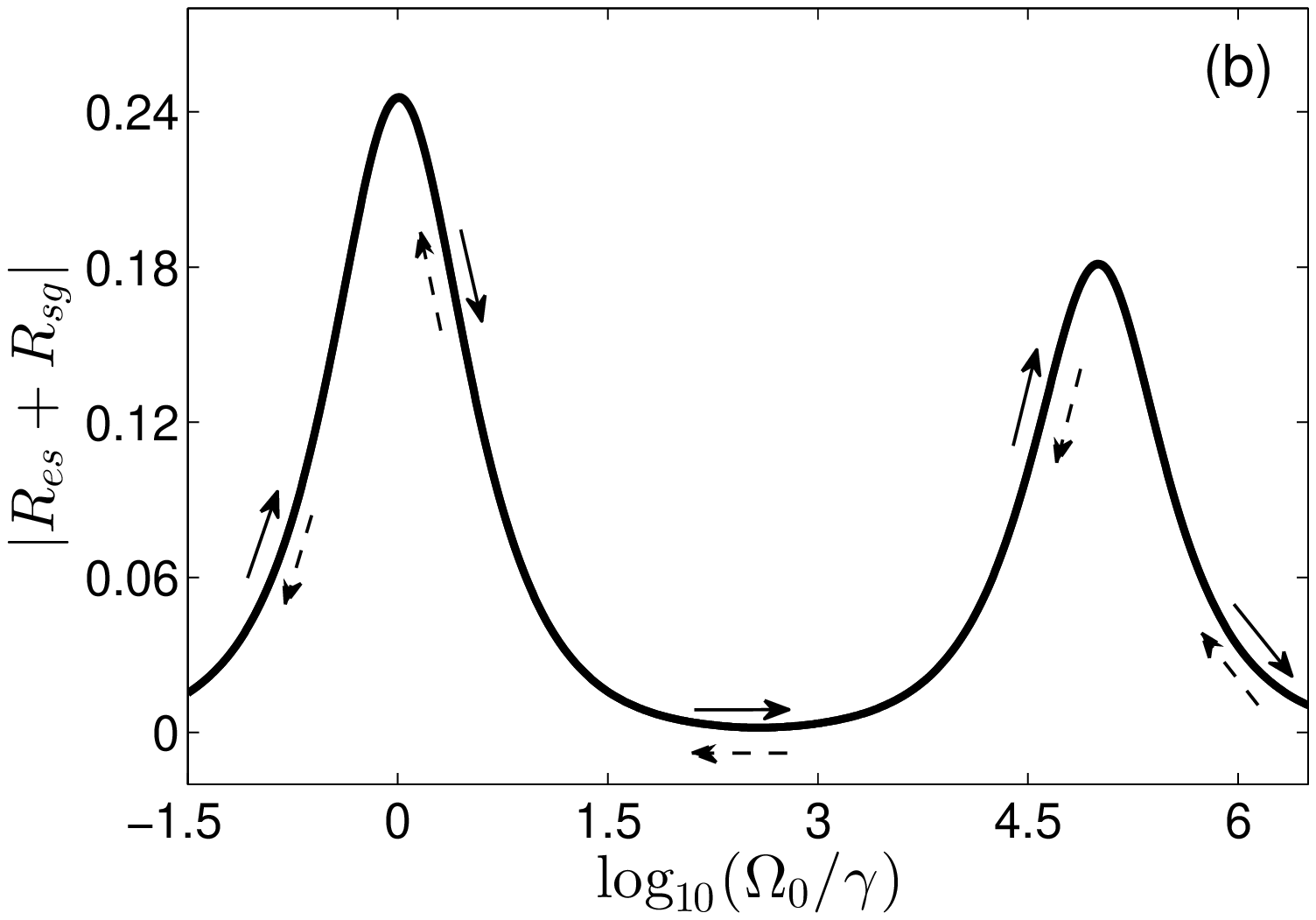}
\end{center}
\caption{Field dependent response of an isolated dimer: (a)~the populations $\rho_{gg}, \rho_{ss}$, and $\rho_{ee}$; (b)~the absolute value of the normalized amplitude of the dimer's dipole moment $|R_{es} + R_{sg}|$. Solid and dashed arrows show the direction of adiabatically sweeping the applied field magnitude $\Omega_0$ up and down, respectively. The frequency of the field is taken to be in resonance with the dimer's bright $|g\rangle \rightarrow |s\rangle$ transition, $\hbar\omega = \hbar\omega_{sg} = \hbar\omega_0 + J = 3.23~\mathrm{eV}$.}
\label{DiagOffDiagVsIntens_Isolated}
\end{figure}

Inspecting Fig.~\ref{DiagOffDiagVsIntens_Isolated}(a), one observes that the populations of an isolated dimer manifest the standard behavior when adiabatically sweeping the applied field strength $\Omega_0$. Let us first discuss the route of increasing field magnitude (indicated in the figure by solid arrows). As expected, in the weak field regime ($\Omega_0 \ll \gamma$), the dimer is in its ground state, $\rho_{gg}\approx 1$ (black curve), while the other states are unpopulated, $\rho_{ss} \approx \rho_{ee} \approx 0$ (red and blue curves). When $\Omega_0$ increases, population starts to transfer from the ground state to the symmetric one-exciton state: $\rho_{gg}$ decreases, accompanied by the growth of $\rho_{ss}$. This occurs when $\Omega_0$ approaches $\gamma$. The population of the doubly excited state still stays close to zero, because the applied field is not resonant with the $|s \rangle \rightarrow |e \rangle$ transition. Increasing $\Omega_0$ further, saturates the $|g \rangle \rightarrow |s \rangle$ transition, which is evident in Fig.~\ref{DiagOffDiagVsIntens_Isolated}(a) from plateaus in the field dependences at $\rho_{gg} = \rho_{ss} = 1/2$. This intermediate regime is characterized by $\Omega_0 \gg \gamma$, but still $\Omega_0 \ll J/\hbar$. Because of the latter condition, the population of the doubly excited state $|e \rangle \rightarrow |s \rangle$ remains close to zero.

Only when $\Omega_0 \sim J/\hbar$, the field is strong enough to overcome the off-resonance detuning of the $|s \rangle \rightarrow |e \rangle$ transition and the latter state is expected to get populated. In fact, at this field magnitude, also the transition $|g \rangle \rightarrow |e \rangle$ takes place via simultaneous absorption of two photons without the creation of population in the s-state as intermediate. In our case, $J/\hbar = 10^5~\gamma$ and, indeed, just at $\Omega_0 \approx 10^{5}\gamma$, the population $\rho_{ee}$ begins to grow, depleting the populations of the symmetric and ground states. In the limit of very high fields, $\Omega_0 \gg J/\hbar$, the transitions between all states are equally probable and the system becomes fully saturated: the populations of all states are the same, $\rho_{gg} = \rho_{ee} = \rho_{ee} = 1/3$. When sweeping down the driving field (indicated in the figure by dashed arrows), all the populations trace back their paths, not showing hysteresis or other deviations (cf. Sec.~\ref{Hybrid}).

Figure~\ref{DiagOffDiagVsIntens_Isolated}(b) presents the field dependence of the absolute value of the normalized amplitude of the dimer's dipole moment $|R_{es} + R_{sg}|$, obtained while adiabatically sweeping $\Omega_0$ up and down, in the same way as above. As is observed, $|R_{es} + R_{sg}|$ shows two peaks and three regions of values close to zero. These observations have clear explanations that can be inferred from the field behavior of the populations, Fig.~\ref{DiagOffDiagVsIntens_Isolated}(a). For $\Omega_0 \ll \gamma$ the depletion of the dimer's ground state is minor and the populations of the excited states $|s \rangle$ and $|e \rangle$ are small. Hence, also the coherences $R_{sg}$ and $R_{es}$ are small, the last one in particular. The first maximum of $|R_{es} + R_{sg}|$ is reached at $\Omega_0 \sim \gamma$, i.e., at the field magnitude for which the population transfer from $|g \rangle \rightarrow |s \rangle$ is developed well already. This causes the coherence $R_{sg}$ to grow in magnitude and arrive at its maximum value for $\Omega_0 \sim \gamma$. At these field magnitudes, the coherence $R_{es}$ is still almost zero, because the $|s \rangle \rightarrow |e \rangle$ transition does not occur yet. In other words, the low-field peak of $|R_{es} + R_{sg}|$ is mainly due to the coherence $R_{sg}$.

When approaching the intermediate saturation regime ($\gamma \ll \Omega_0 \ll J/\hbar$), the populations $\rho_{ss}$ and $\rho_{gg}$ tend to the same value of 1/2. Because the coherence $R_{sg}$ is proportional to $\rho_{ss} - \rho_{gg}$, it drops to zero in this regime; as $R_{es}$ still is negligible, also $|R_{es} + R_{sg}|$ drops to zero. In the strong saturation regime, when $\Omega_0 > J/\hbar$, the transition $|s \rangle \rightarrow |e \rangle$ occurs and the coherence $R_{es}$ starts to grow efficiently, giving rise to the second peak of $|R_{es} + R_{sg}|$. It should be noticed that the coherence $R_{sg}$ also is not zero now, which explains the difference in amplitudes of the two peaks. Upon further increasing the driving field magnitude, all transitions of the dimer approach the truly saturated regime, whence both $R_{sg}$ and $R_{es}$ decrease, and thus $|R_{es} + R_{sg}|$ drops to zero as well. Again, sweeping back, no hysteresis is observed. 

\subsection{Dimer-MNP hybrid}
\label{Hybrid}

\begin{figure}[ht]
\begin{center}
\includegraphics[width=0.80\linewidth]{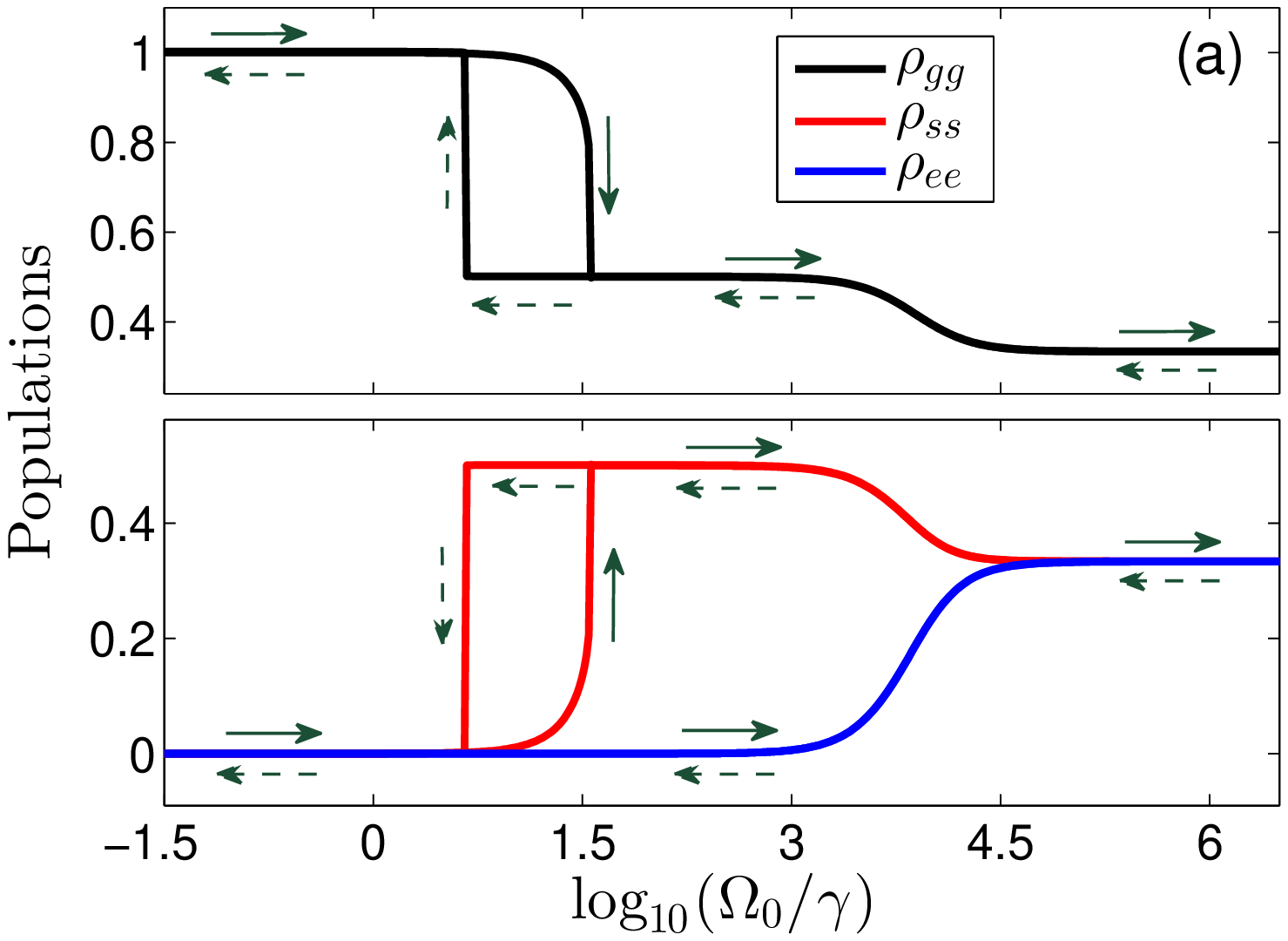}
\includegraphics[width=0.80\linewidth]{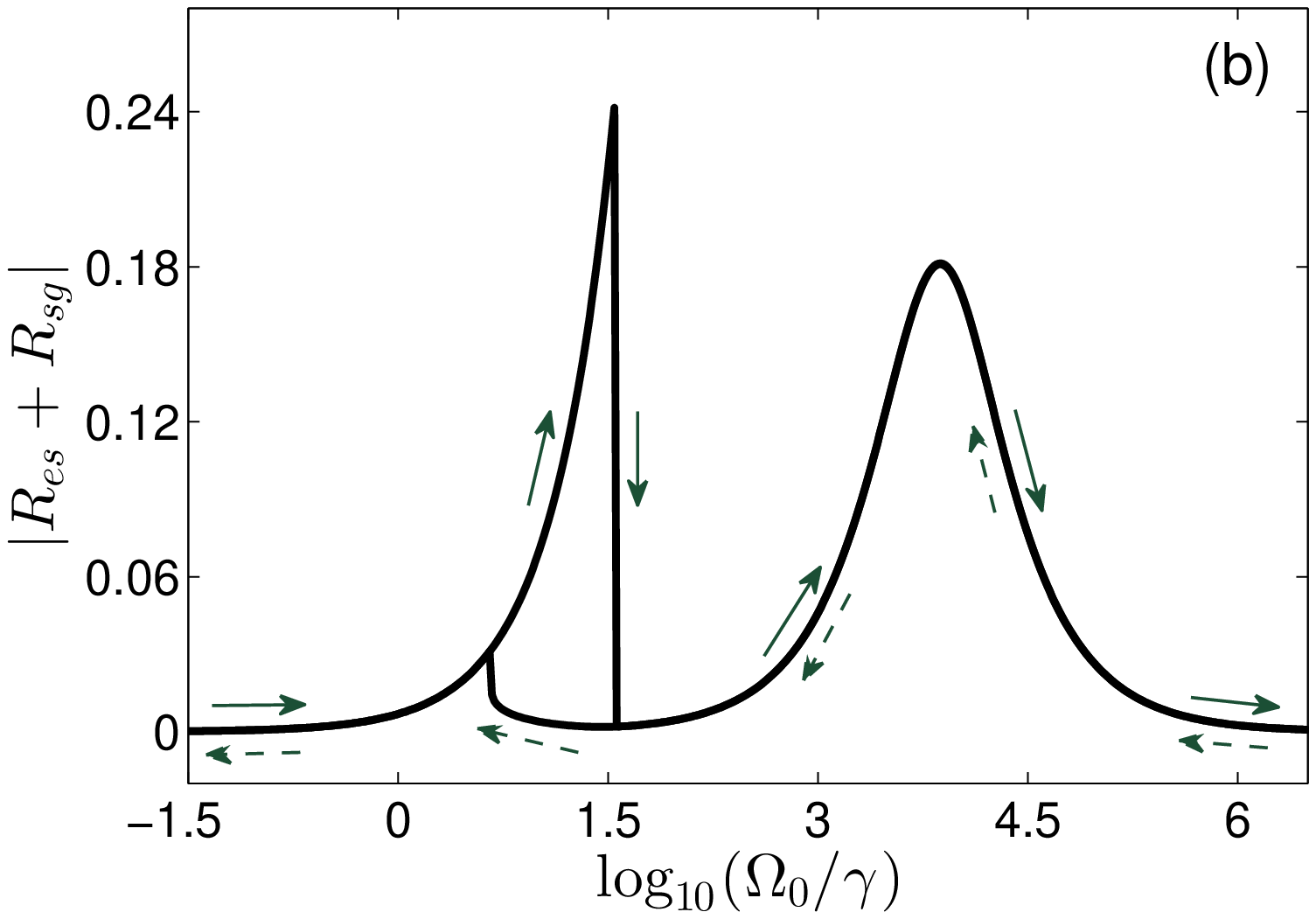}
\end{center}
\caption{Same as in Fig.~\ref{DiagOffDiagVsIntens_Isolated}, but now for a dimer-MNP hybrid with the centre-to-centre distance between the MNP and the dimer of $d=18~\mathrm{nm}$. (a) ~ The populations $\rho_{gg}$, $\rho_{ss}$, and $\rho_{ee}$. (b)~ Absolute value of the normalized amplitude of dimer's dipole moment $|R_{es} + R_{sg}|$.}
\label{DiagOffDiagVsIntens_Coupled}
\end{figure}

The optical response of a dimer-MNP hybrid is shown in Fig.~\ref{DiagOffDiagVsIntens_Coupled}. Here, we set the center-to-center distance between the MNP and the dimer at $d=18~\mathrm{nm}$; the other parameters are as in Sec.~\ref{Isolated dimer}. Comparing this figure to Fig.~\ref{DiagOffDiagVsIntens_Isolated}, it is obvious that the hybridization leads to strong changes of the response. In contrast to the isolated dimer, the response of the hybrid depends on the history of sweeping the applied field (up or down): in a certain interval of $\Omega_0$, the quantities $\rho_{gg}$, $\rho_{ss}$, and $|R_{es} + R_{sg}|$ exhibit a hysteresis loop, related to the $|g \rangle \rightarrow |s \rangle$ transition. This indicates bistability of the hybrid.~\cite{gibbs1985optical} The effect results from the self-action of the dimer via the MNP, described by the feed-back parameter $G$ [Eq.~(\ref{G})] and also occurs for a hybrid of a MNP and a monomer~\citep{MalyshevPRB2011, NugrohoJCP2013}, as it only requires one exciton transition.

The real and imaginary  parts of $G$ (denoted $G_R$ and $R_I$, respectively), give rise to different mechanisms for bistability, as has been analyzed in detail in Ref.~\onlinecite{NugrohoJCP2013}. In order to achieve bistable response, the values of $G_R$ and $G_I$ should meet certain threshold conditions.~\cite{MalyshevPRA2012,NugrohoJCP2013} For the parameters chosen in Fig.~\ref{DiagOffDiagVsIntens_Coupled}, we have $G = (27.47 + 936.64~i)\Gamma$ [$\hbar G = (0.05 + 1.83~i)\mathrm{meV}$], and indeed the criteria for bistability are met.

From comparing Figs.~(\ref{DiagOffDiagVsIntens_Isolated})(a) and (\ref{DiagOffDiagVsIntens_Coupled})(a), one observes that the field magnitudes at which the two saturation steps occur differ. For the hybrid the saturation of the one-exciton transition, indicated by the plateau at $\rho_{ee}=\rho_{ss}=1/2$ occurs at a value of $\Omega_0$ that is larger than for the isolated dimer. The explanation lies in the (population dependent) renormalization of the resonance frequency and the coherence relaxation rate due to the dimer-MNP coupling: $\omega_{sg} \rightarrow \omega_{sg}+G_R Z_{sg}$ and $\Gamma \rightarrow \Gamma - G_I Z_{sg}$. At low values of $\Omega_0$, the population difference $Z_{sg}=-1$, so that the renormalization brings the effective resonance of the dimer away from the frequency of the applied field and at the same time increases the effect of damping. Realizing that both $G_R$ and $G_I$ are much larger than $\Gamma$, these renormalizations imply that a higher value of $\Omega_0$ is required to saturate the transition. Oppositely, for the hybrid, the two-exciton transition saturates at a lower intensity than for the isolated dimer. The explanation lies in the enhancement of the field acting on the dimer due to the exciton plasmon interaction [see Eq.~(\ref{OmegaD})]. All the above effects also manifest themselves in shifts of the peaks in the field dependencies of $|R_{es} + R_{sg}|$ [Figs.~(\ref{DiagOffDiagVsIntens_Isolated})(b) and (\ref{DiagOffDiagVsIntens_Coupled})(b)].

\begin{figure}[ht]
\begin{center}
\includegraphics[width=0.80\columnwidth]{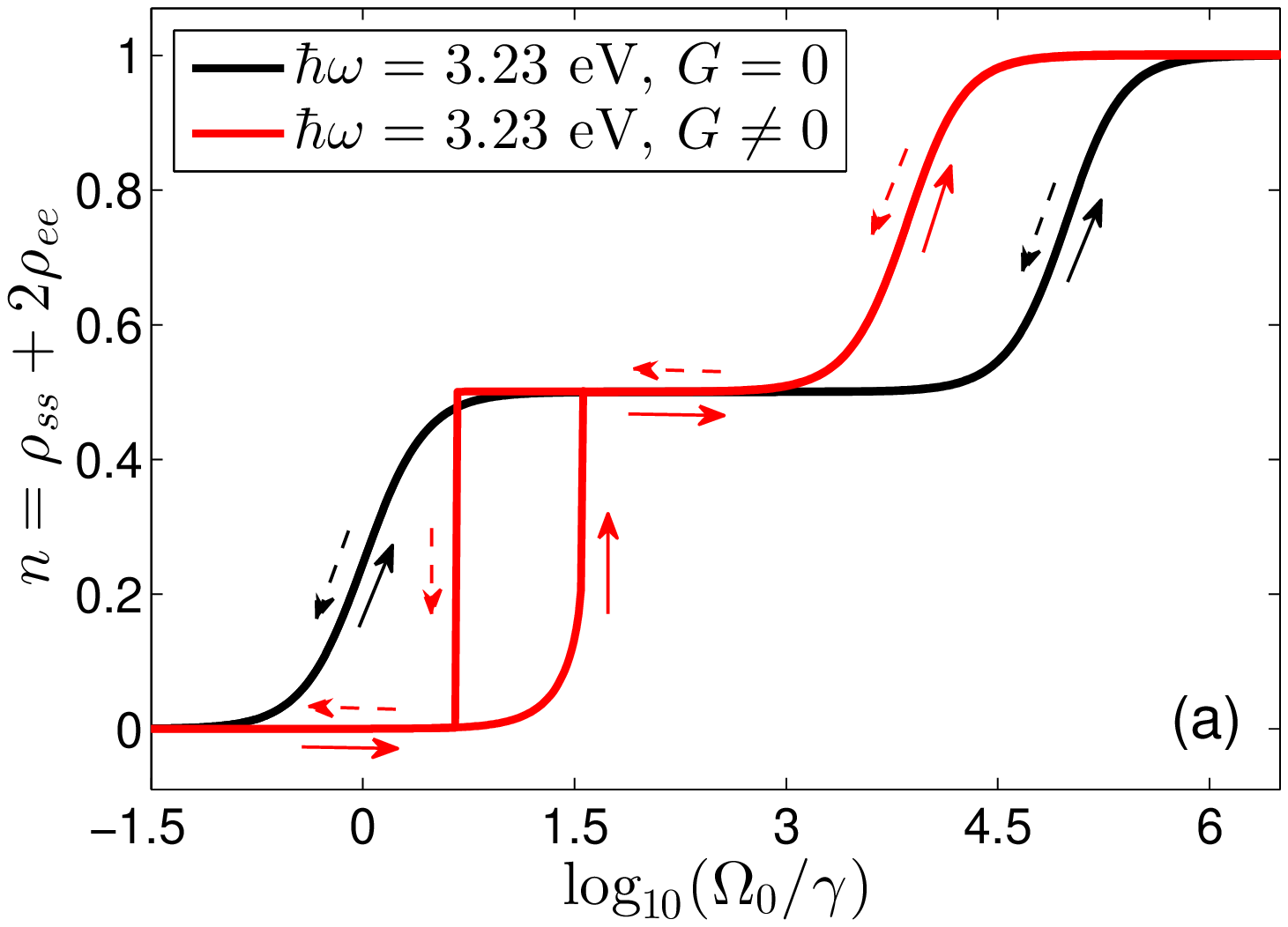}
\includegraphics[width=0.80\columnwidth]{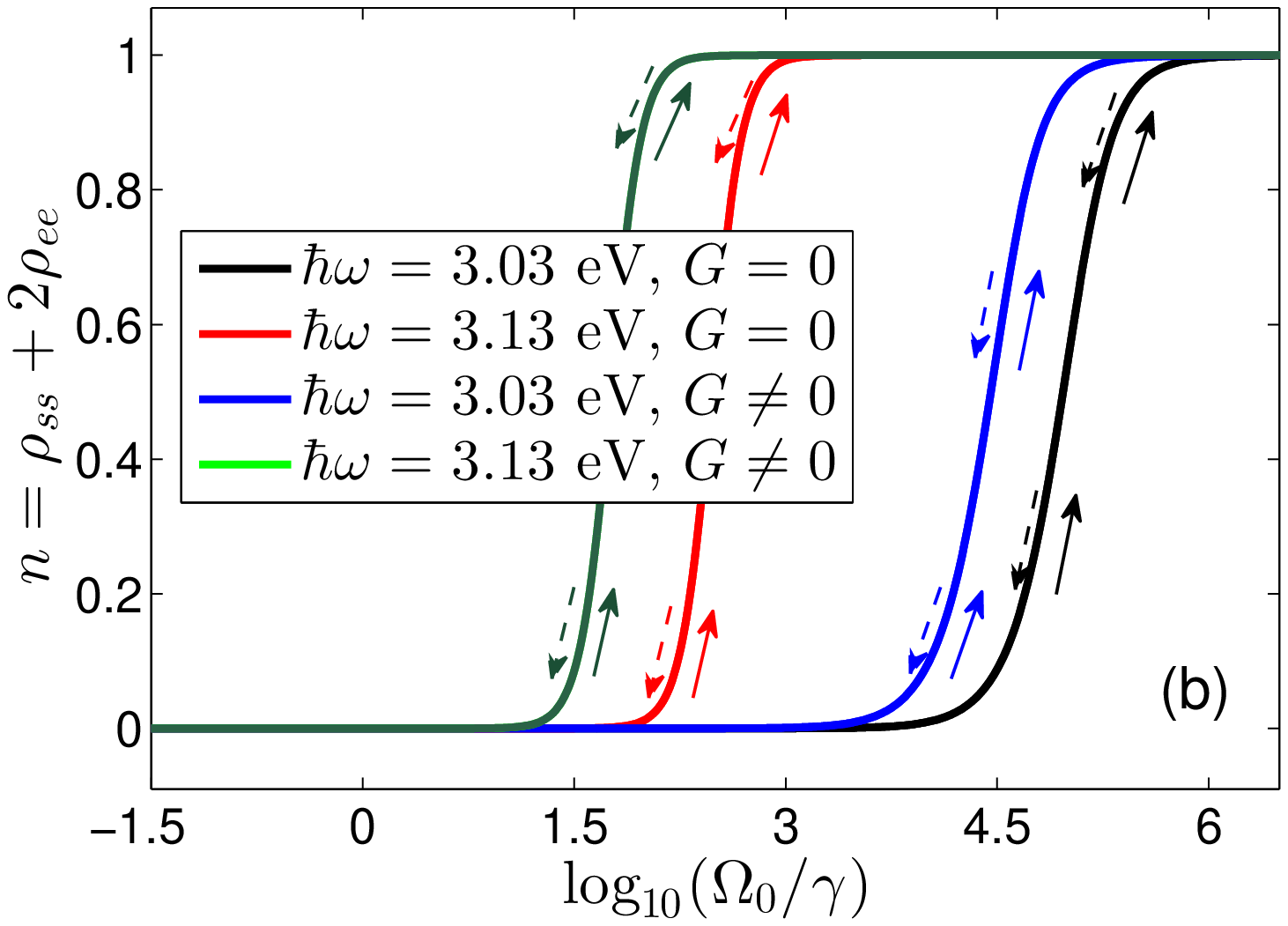}
\end{center}
\caption{Field dependence of the average number of excited emitters, $n = \rho_{ss}+2\rho_{ee}$, for an isolated dimer ($G=0$) and for a dimer coupled to a MNP ($G \neq 0$),  calculated for three types of excitation: (a) ~ in resonance with the $|g\rangle \rightarrow |s\rangle$ transition, $\hbar\omega = \hbar\omega_{sg} = \hbar\omega_0 + J = 3.23~\mathrm{eV}$, (b) ~ in resonance with the two-photon transition $|g\rangle \rightarrow |e\rangle$, with frequency $\hbar\omega = \hbar\omega_0 = 3.13~\mathrm{eV}$, and in resonance with the $|s\rangle \rightarrow |e\rangle$ transition, $\hbar\omega = \hbar\omega_{0}-J = 3.03\mathrm{eV}$. As before, the pure dephasing constant equals $\Gamma^{\prime} = 1.08\gamma$. Solid and dashed arrows show the direction of adiabatically sweeping the applied field magnitude $\Omega_0$.}
\label{IO}
\end{figure}

To get more insight into the field dependence of the hybrid's response, we calculated the average number of excited emitters, given by $n = \rho_{ss} + 2\rho_{ee}$, for three different frequencies of the applied field: $\hbar\omega = \hbar\omega_{sg} = \hbar\omega_{0} + J = 3.23~\mathrm{eV}$ (in resonance with the $|g\rangle \rightarrow |s\rangle$ transition as well as the MNP plasmon $\omega_{sp} = 3.23~\mathrm{eV}$), $\hbar\omega = \hbar\omega_{0} = 3.13~\mathrm{eV}$ (in resonance with the two-photon transition $|g\rangle \rightarrow |e\rangle$), and $\hbar\omega = \hbar\omega_{es} = \hbar\omega_{0}-J = 3.03~\mathrm{eV}$ (in resonance with the $|s\rangle \rightarrow |e\rangle$ transition). Figure~\ref{IO} shows the results for an isolated dimer ($G=0$) and a dimer-MNP hybrid ($G \neq 0$), adiabatically sweeping the applied field magnitude $\Omega_0$ up and down.

For $\hbar\omega = \hbar\omega_{sg} =  3.23~\mathrm{eV}$ [Fig~\ref{IO}(a)], $n$ clearly displays bistable behavior in the hybrid. Upon sweeping $\Omega_0$ up, initially $n$ remains close to zero, meaning that the dimer is in its ground state $|g \rangle$. At some critical value of $\Omega_0 \approx 10^{1.5}\gamma$, the $|g\rangle \rightarrow |s\rangle$ transition, dissimilar to an isolated dimer, is abruptly saturated ($n = 1/2$), maintaining this state within some interval of $\Omega_0$. Upon increasing $\Omega_0$ beyond that interval, all dimer transitions become saturated. Sweeping $\Omega_0$ down again, the transition from the fully saturated regime ($n=1$) to the intermediate one ($n=1/2$) shows no hysteresis, while the second step down to the unsaturated regime ($n = 0$) again occurs abruptly, but at a lower value of $\Omega_0$ than the upward step. Thus, the average number of excited emitters exhibits hysteresis, confirming that the dimer coupled to the MNP is a bistable system, where the bistability occurs only in the $|g\rangle \rightarrow |s\rangle$ channel and does not appear in the $|s\rangle \rightarrow |e\rangle$ channel. No multistable~\cite{AsadpourOC2014} behaviour is found.

If the driving field is tuned away from the $|g\rangle \rightarrow |s\rangle$ resonance, Fig.~\ref{IO}(b), no bistability occurs in the hybrid's optical response. However, the effect of local-field enhancement due to the plasmon excitation is seen for all cases: the applied field magnitude $\Omega_0$ required to completely saturate the system always reduces significantly in the presence of the MNP. We finally note that for the field frequency $\hbar\omega = \hbar\omega_0 = 3.13 \mathrm{eV}$, the system gets fully saturated at the lowest intensity as compared to all others excitation conditions. This can be understood from the fact that then the applied field is in exact resonance with the $|g\rangle \rightarrow |e\rangle$ two-photon transition.

\section{Hybrid absorption spectra}
\label{Spect}

In this section, we analyse the absorption spectrum of the dimer-MNP hybrid and show that its shape strongly depends on the detuning between the one-exciton resonance $\omega_{sg}$ and the plasmon resonance $\omega_{sp}$, as well as the ratio $|G|/\Gamma$. We will distinguish between the weak-field limit $\Omega_0 \approx \gamma$, where the two-exciton state $|e\rangle$ is irrelevant, and strong-field limit ($\Omega_0 \geq 10^2\gamma$), where the state $|e\rangle$ participates in the process.

The spectra are computed from the time average of the power (energy per second) $Q$ absorbed by the hybrid; $Q$ consists of two contributions, one from the dimer ($Q_\mathrm{DIM}$) and the other from the MNP  ($Q_\mathrm{MNP}$): $Q = Q_\mathrm{DIM} + Q_\mathrm{MNP}$. From now on, we will refer to the frequency dependence of all $Q$'s as the absorption spectra of the corresponding parts.

The absorption of the dimer results from the creation of excitons, while in the case of the MNP, it is due to charge oscillations (plasmons). Here, we do not include the contribution of scattering by the MNP, because for small particles ($r<<\lambda$) this effect is negligible. The power absorbed by the dimer (MNP) is given by:~\cite{Yariv1989}:
\begin{equation}
  Q_\mathrm{DIM(MNP)} = \frac{1}{2} \varepsilon_0\omega V_{\mathrm{DIM(MNP)}}
                        \mathrm{Im}[\chi_\mathrm{DIM(MNP)}]|E_0|^2~,
\label{Q_DIM(MNP)}
\end{equation}
where $\chi_\mathrm{DIM(MNP)}$ is the susceptibility of the dimer (MNP), defined as the proportionality coefficient between the  polarization (dipole moment per unit volume) and the applied field amplitude $E_0$, and $V_\mathrm{DIM(MNP)}$ is the volume of the dimer (MNP).

The dimer susceptibility $\chi_\mathrm{DIM}$ is calculated quantum mechanically and has the form
\begin{equation}
  \chi_\mathrm{DIM} = \frac{D(R_{es}+R_{sg})}
                      {2 V_\mathrm{DIM} \varepsilon_0  E_\mathrm{0}} \ .
\label{chiDIM}
\end{equation}
Note that the volume $V_{DIM}$ attributed to the dimer is somewhat arbitrary. However, in the absorbed power, Eq.~(\ref{Q_DIM(MNP)}), this volume drops out so that this arbitrariness is irrelevant.

The MNP susceptibility $\chi_\mathrm{MNP}$ is computed within the framework of classical electrodynamics:
\begin{equation}
  \chi_\mathrm{MNP} = \frac{\varepsilon_{b}\alpha}{V_{MNP}}
                      \left[ 1
                    + \frac{D}{\pi \varepsilon_0 \varepsilon_b d^3 E_0}
                      (R_{es}+R_{sg}) \right]~,
\label{chiMNP}
\end{equation}
The second term in square brackets represents a correction to the external field $E_0$ by the field produced by the dimer at the position of the MNP. Having obtained $\chi_\mathrm{DIM}$ and $\chi_\mathrm{MNP}$, we can now calculate the absorption spectrum $Q$ of the dimer-MNP composite.

\subsection{Weak field regime}
\label{LinSpect}

\begin{figure}[h!]
\centering
\includegraphics[width=0.49\linewidth]{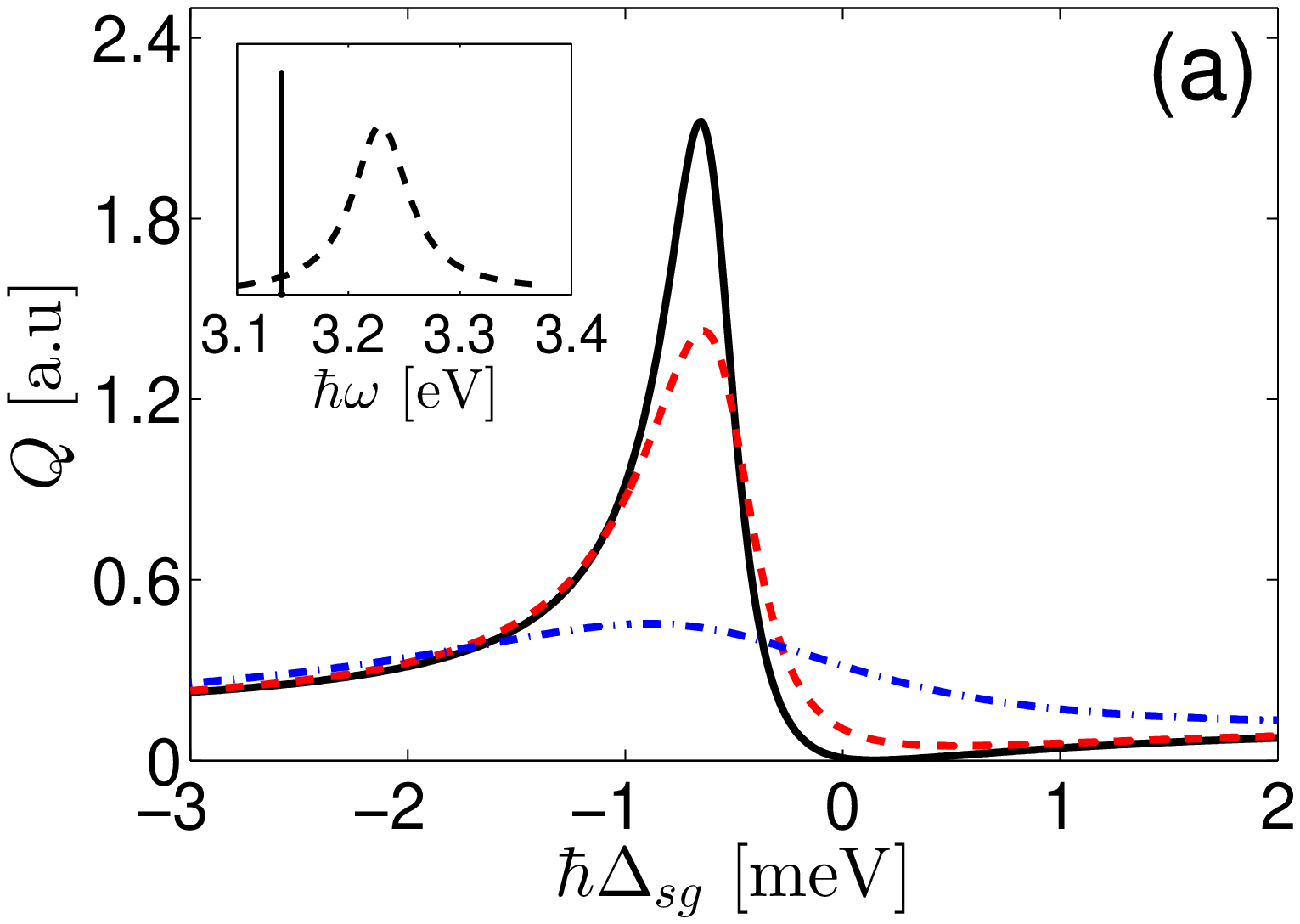}
\includegraphics[width=0.49\linewidth]{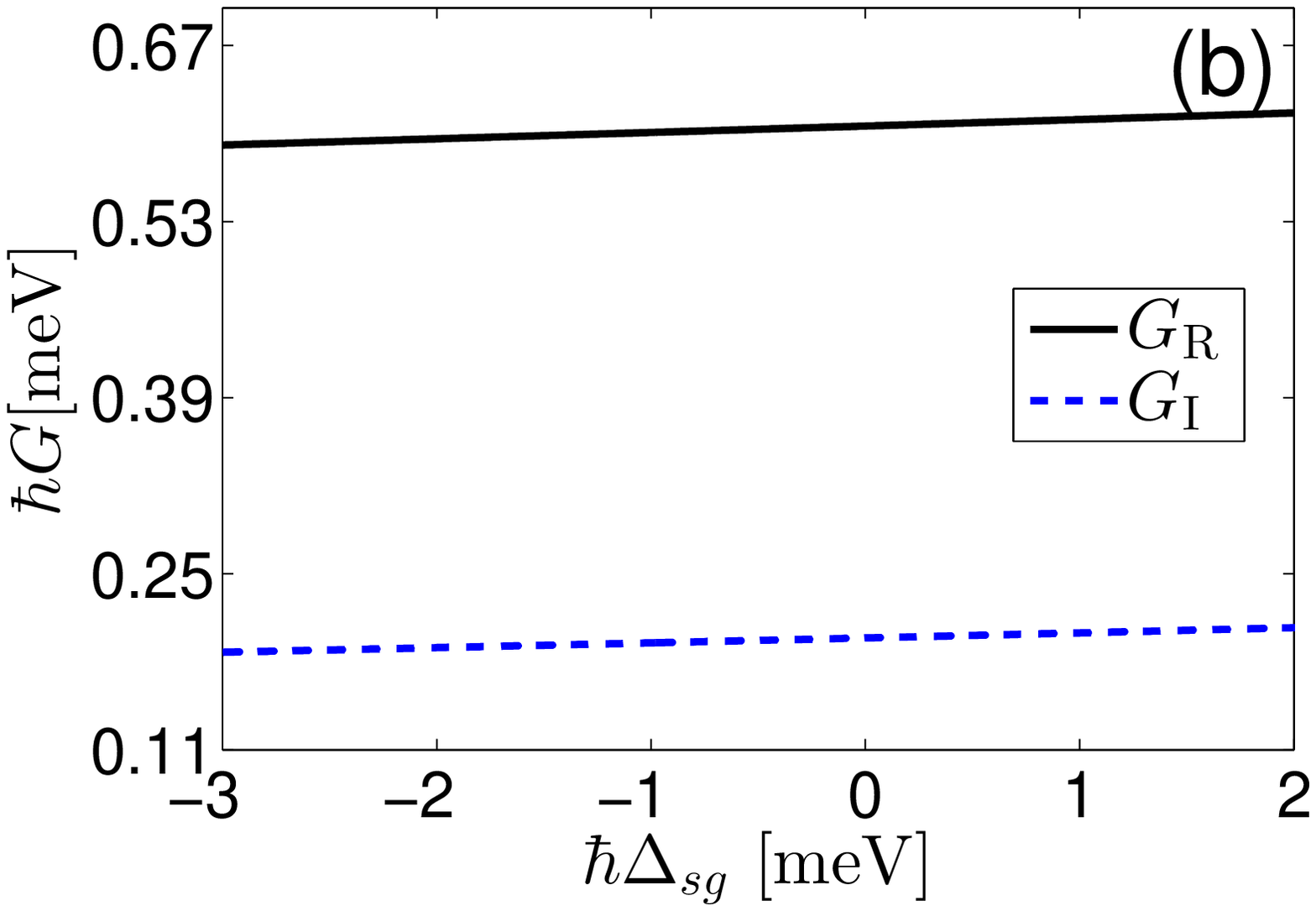}
\includegraphics[width=0.49\linewidth]{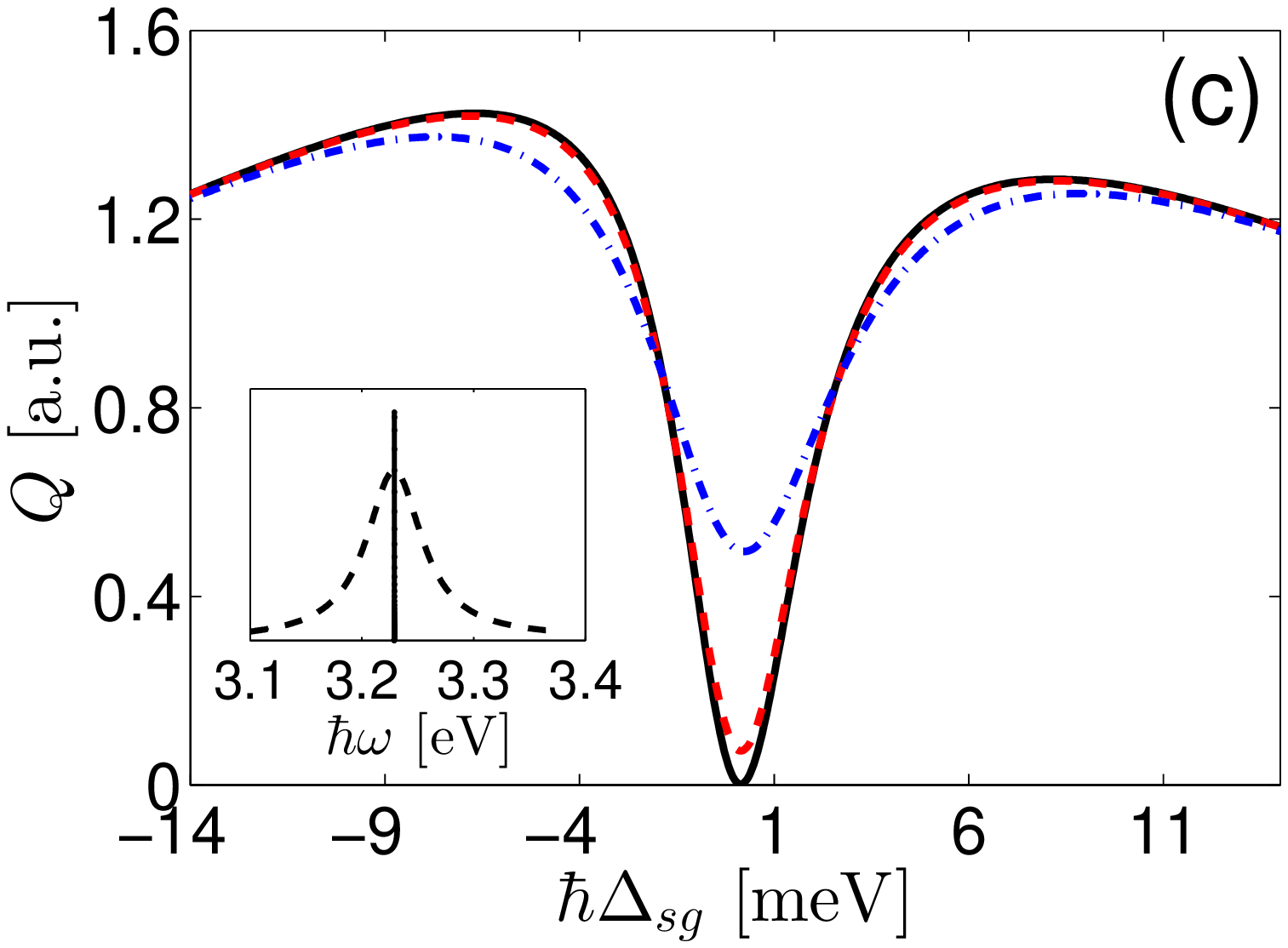}
\includegraphics[width=0.49\linewidth]{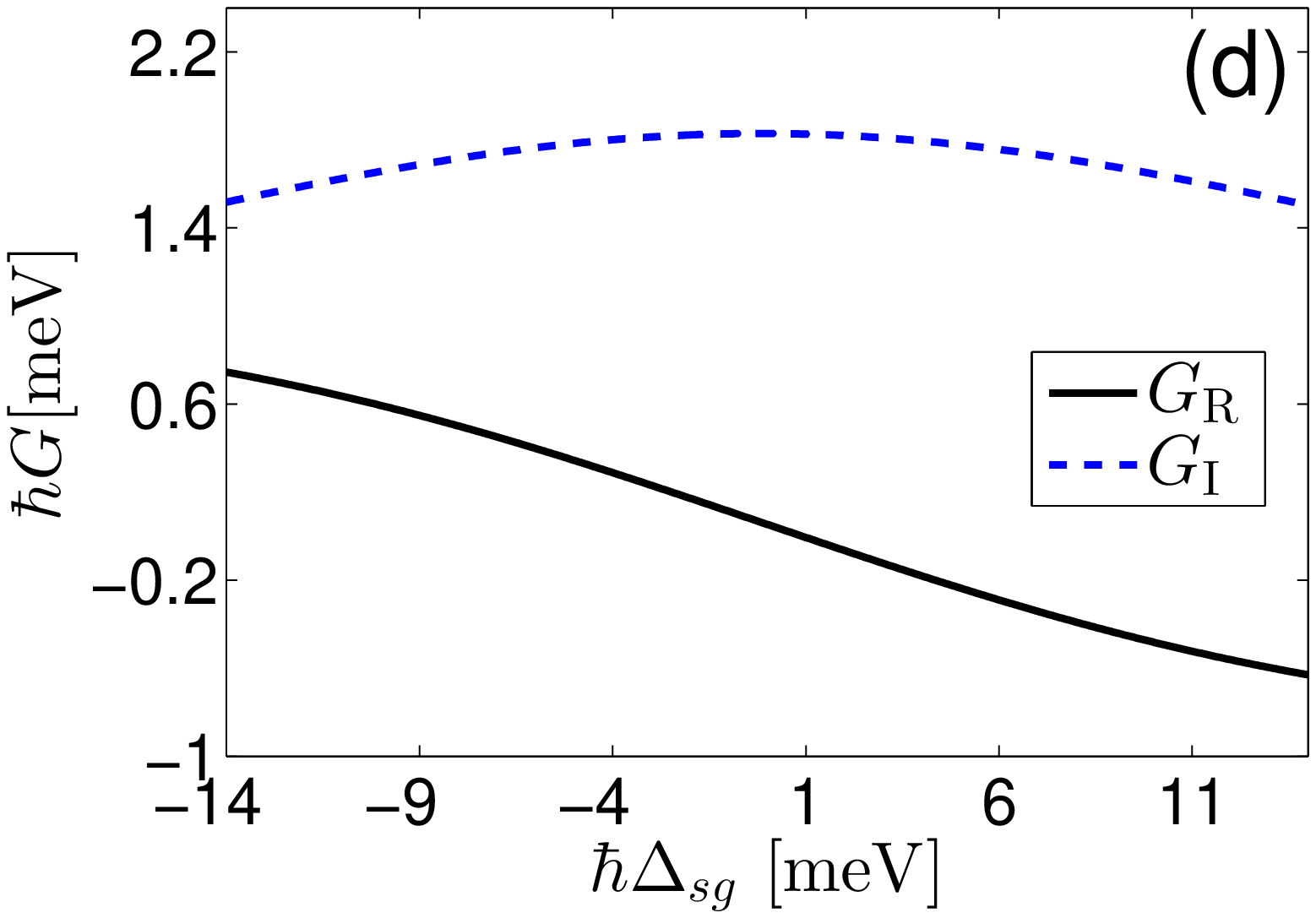}
\includegraphics[width=0.49\linewidth]{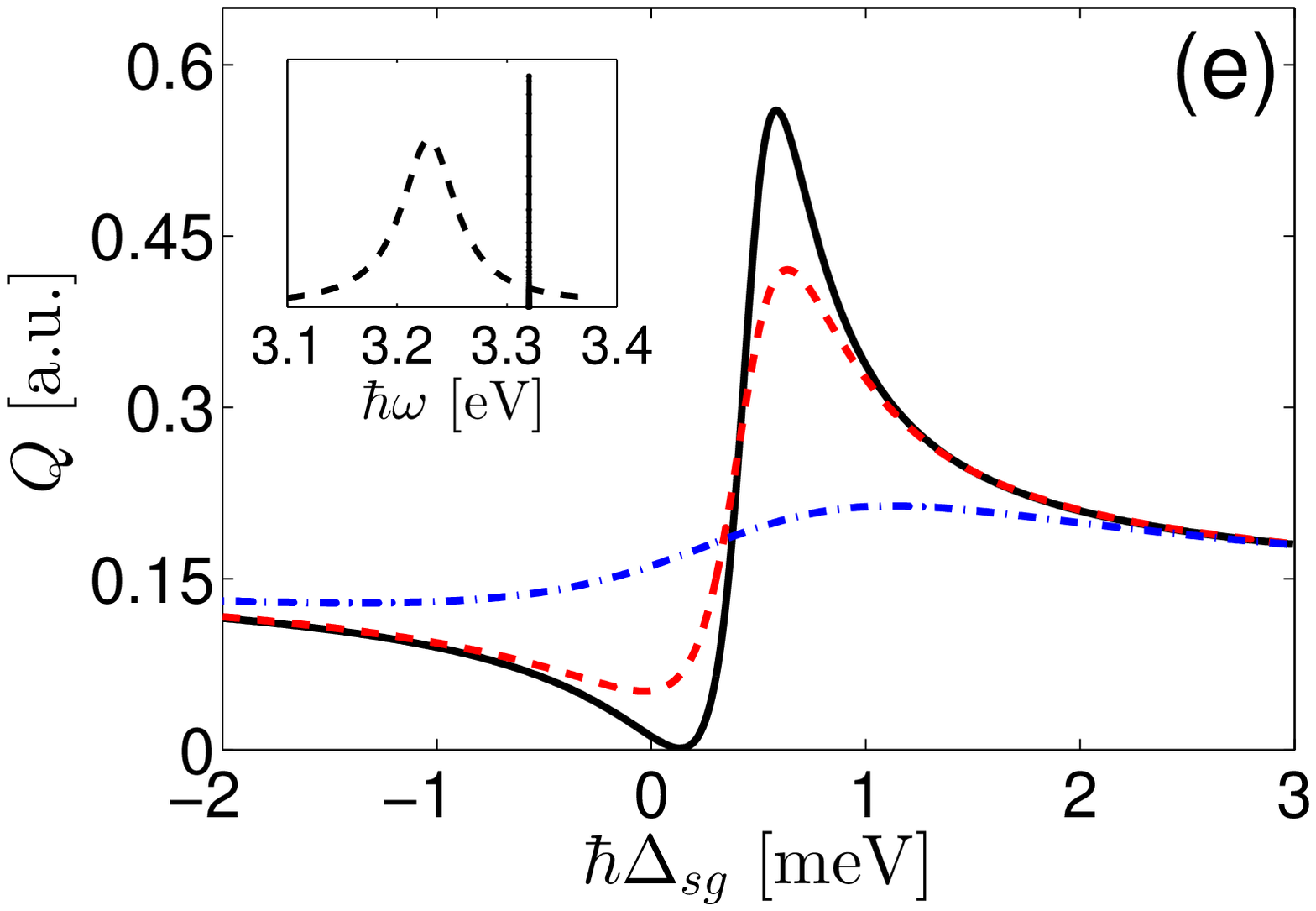}
\includegraphics[width=0.49\linewidth]{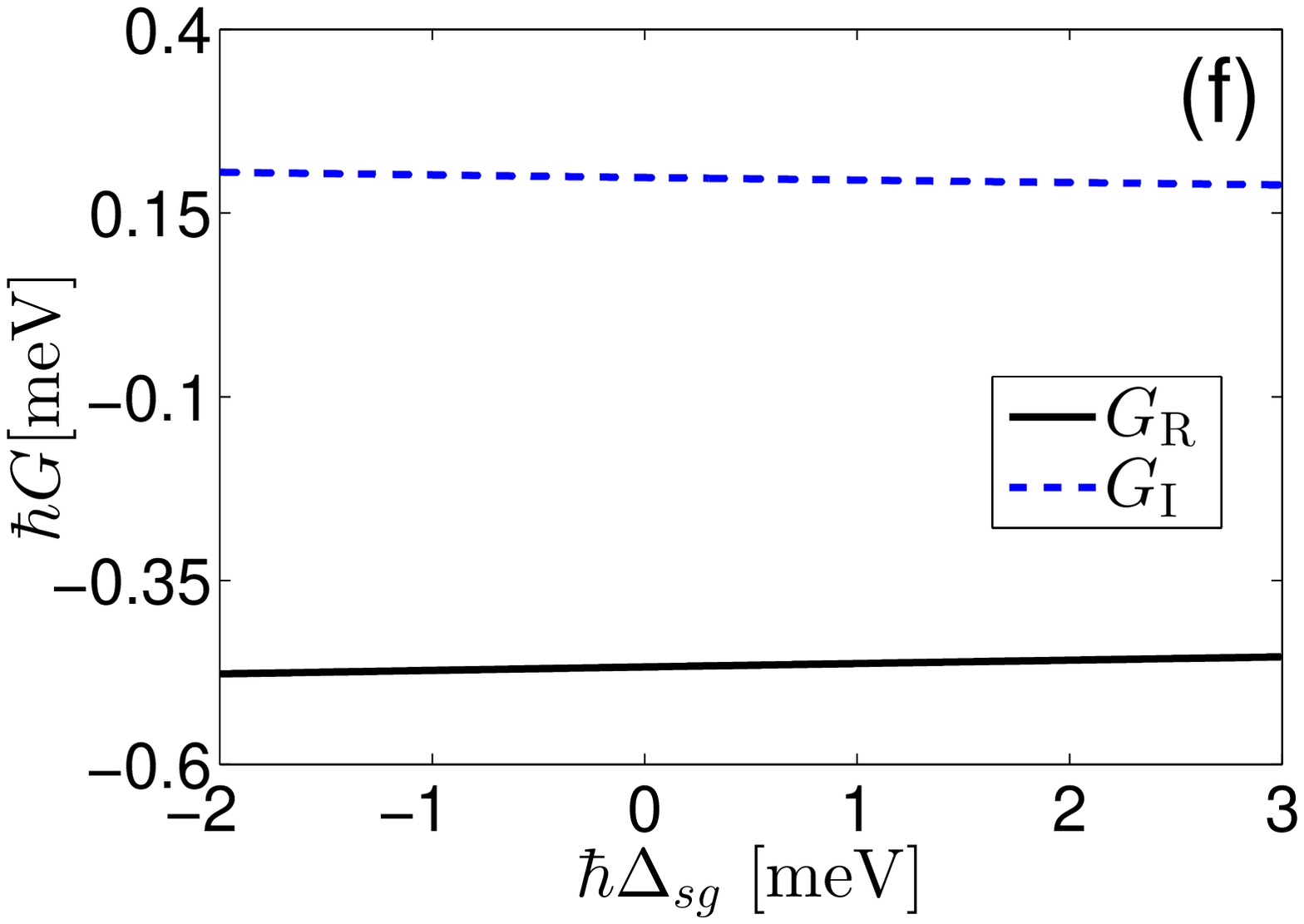}
\caption{Absorption spectra of the dimer-MNP hybrid calculated for different locations of the isolated dimer resonance $\omega_{sg}$ with respect to the MNP plasmon resonance $\omega_{sp} = 3.23~\mathrm{eV}$. (a)~$\hbar\omega_{sg} = 3.14~\mathrm{eV}$ (below the plasmon resonance), (c)~$\hbar\omega_{sg} = 3.23~\mathrm{eV}$ (at the plasmon resonance), and (e)~$\hbar\omega_0=3.32~\mathrm{eV}$ (above the plasmon resonance). The inserts illustrate these three positions: the dashed curves represent the MNP plasmon peak, while the vertical lines indicate the isolated dimer resonance. Each panel displays the spectra of the hybrid computed for different dephasing rates: $\Gamma^{\prime} = 1.08 ~\gamma$ ($\hbar\Gamma^{\prime} = 10^{-3}~\mathrm{meV}$) (black solid curve), $\Gamma^{\prime} = 1.08\times 10^2~\gamma$ ($\hbar\Gamma^{\prime} = 0.1~\mathrm{meV}$) (red dashed curve), and $\Gamma^{\prime} = 1.08\times 10^3\gamma$ ($\hbar\Gamma^{\prime} = 1~\mathrm{meV}$) (blue dash-dotted curve). Panels (b), (d), and (f) show the spectra of the real and imaginary parts of the coupling parameter $G$.}
\label{SpectLinearFano}
\end{figure}

First, we address the linear absorption spectrum of a dimer-MNP hybrid under weak field excitation. In particular, we set the Rabi frequency $\Omega_0 = \gamma$. Then, the $|s \rangle \rightarrow |e \rangle$ transition is not involved in the hybrid's absorption. We consider different positions of the frequency of the dimer's one-exciton resonance $\omega_{sg}$ with respect to the plasmon resonance frequency $\omega_{sp}$ and vary the dephasing rate $\Gamma^{\prime}$. The other parameters used in the calculations are identical to those described above.

Figure~\ref{SpectLinearFano} shows the absorption spectra obtained for a dimer-MNP hybrid with center-to-center distance $d = 18~\mathrm{nm}$ (strong coupling regime) at three values of the one-exciton resonance: $\hbar\omega_{sg} = 3.14~\mathrm{eV}$ (panel (a), below the plasmon resonance), $\hbar\omega_{sg} = 3.23~\mathrm{eV}$ (panel (c), at the plasmon resonance), and $\hbar\omega_{sg} = 3.32~\mathrm{eV}$ (panel (e), above the plasmon resonance). In each case, results are shown calculated for three different dephasing rates:  $\Gamma^{\prime} = 1.08~\gamma$ \, ($\hbar\Gamma^{\prime} = 10^{-3}~\mathrm{meV}$), $\Gamma^{\prime} = 1.08\times10^2~\gamma$ \, ($\hbar\Gamma^{\prime} = 0.1~\mathrm{meV}$), and $\Gamma^{\prime} = 1.08\times10^3~\gamma$ \, ($\hbar\Gamma^{\prime} = 1~\mathrm{meV}$). In the absence of the exciton-plasmon interaction, the spectrum would be a simple Lorentzian lineshape centered at $\Delta_{sg}=0$ and with a width given by $\gamma + \Gamma^{\prime}$, superimposed on the broad Lorentzian plasmon peak centered at $\Delta_{sg} = \omega_{sp}-\omega_{sg}$. As is clearly seen, the exciton-plasmon interaction, given by the coupling parameter $G$, strongly affects the spectrum, which now exhibits an asymmetric dispersive shape with a dip to almost zero absorption (for small $\Gamma^{\prime}$ values), resembling the well-known Fano resonance~\citep{FanoPhysRev1961}. Recall that the essence of the original quantum Fano effect - response of a system comprising a discrete level interacting with a dark quantum continuum - consists of the quantum interference of the mixing amplitudes from the bare states to the hybridized (due to the interaction) collective states. In our case, the continuum is a classical system with losses: surface plasmon of the MNP. Here, the quantum picture can not be directly applied, but by analogy, the interference of the applied field and the field produced by the hybrid
leads to the Fano-like spectrum.~\cite{ZhangPRL2006,ArtusoNanoLett2008} Whether the interference is constructive (peak) or destructive (dip) depends on the detuning away from the exciton resonance.

Panels (b), (d), and (f) in Fig.\ref{SpectLinearFano} show the frequency dependence of the coupling parameter $G$ in the relevant frequency ranges. Evidently, when the exciton resonance lies in one of the tails of the plasmon peak, both $G_R$ and $G_I$ are almost constant [Figs.~\ref{SpectLinearFano}(b) and (f)]. However, when the exciton is (close to) resonant with the plasmon, $G$ (in particular $G_R$) shows a pronounced frequency dependence [Fig.~\ref{SpectLinearFano}(d)]. Note that, when $\omega_{sg}$ lies in the low-frequency tail of the plasmon peak, the real part of $\alpha(\omega)$  is positive and therefore $G_R$ is positive as well [Fig.\ref{SpectLinearFano}(b)]. The peak in the spectrum then occurs at lower frequency then the dip [Fig.\ref{SpectLinearFano}(a)]. For $\omega_{sg}$ in the higher frequency tail of the plasmon peak, $G_R$ is negative and the peak in the spectrum occurs at higher frequency than the dip [Figs.~\ref{SpectLinearFano}(e) and (f)].

We will first make several observations regarding the coupled exciton-plasmon system under resonant conditions. Whether or not the dimer-MNP system is coherently coupled, depends, in fact, on the ratio of the dimer-MNP coupling strength $|G|$ and the dephasing rate $\Gamma = \gamma + \Gamma^{\prime}$. To investigate this, consider the situation where the dimer resonance approaches the maximum of the plasmon peak ($\omega_{sg} \rightarrow \omega_{sp}$), for which the spectrum shows only a dip in the absorption and no peak [Fig.~\ref{SpectLinearFano}(c)]. In this case, the imaginary part $\mathrm{Im[\alpha]}$ acquires its maximum, whereas $\mathrm{Re[\alpha]}$ tends to zero [see Fig.~\ref{polarizability}]. The latter allows one to derive a simple analytical formula for the susceptibility of the dimer and the MNP (see Appendix~\ref{ApendixB}) which leads to the totally absorbed power by the hybrid in the form

\begin{equation}
  Q = \frac{1}{4}\omega
      \left(
      \varepsilon_0\varepsilon_b\alpha_I + \frac{D^2}{2\hbar} \frac{1}{\Gamma}
      \right)
      \frac{\Gamma}{\Gamma+G_I} \ .
\label{Qdip}
\end{equation}
Here, $\alpha_I = \mathrm{Im}[\alpha]$. The first and second terms in brackets correspond to the absorption of the MNP and the dimer, respectively. As is seen from Eq.~(\ref{Qdip}), the minimum value of the hybrid absorption (reached in the dip) is governed by the ratio $G_I/\Gamma$. 

The physical explanation is as follows. In the strong coupling regime ($G_I \gg \Gamma$), both the dimer and the MNP acquire a large collective damping, given by $G_I$, thus reducing the hybrid's absorption. The dephasing counteracts the collective nature of the hybrid's states and works towards recovering the absorption near $\omega_{sp}$, which finally, at $G_I \ll \Gamma$, is represented by the absorption of the isolated MNP and dimer (first and second terms in the parentheses of Eq.~\ref{Qdip}, respectively).
This is the origin of the dependence of the dip depth on the ratio $G_I/\Gamma$, reflected in Fig.~\ref{SpectLinearFano}(c).

\begin{figure}[h!]
\centering
\includegraphics[width=0.80\linewidth]{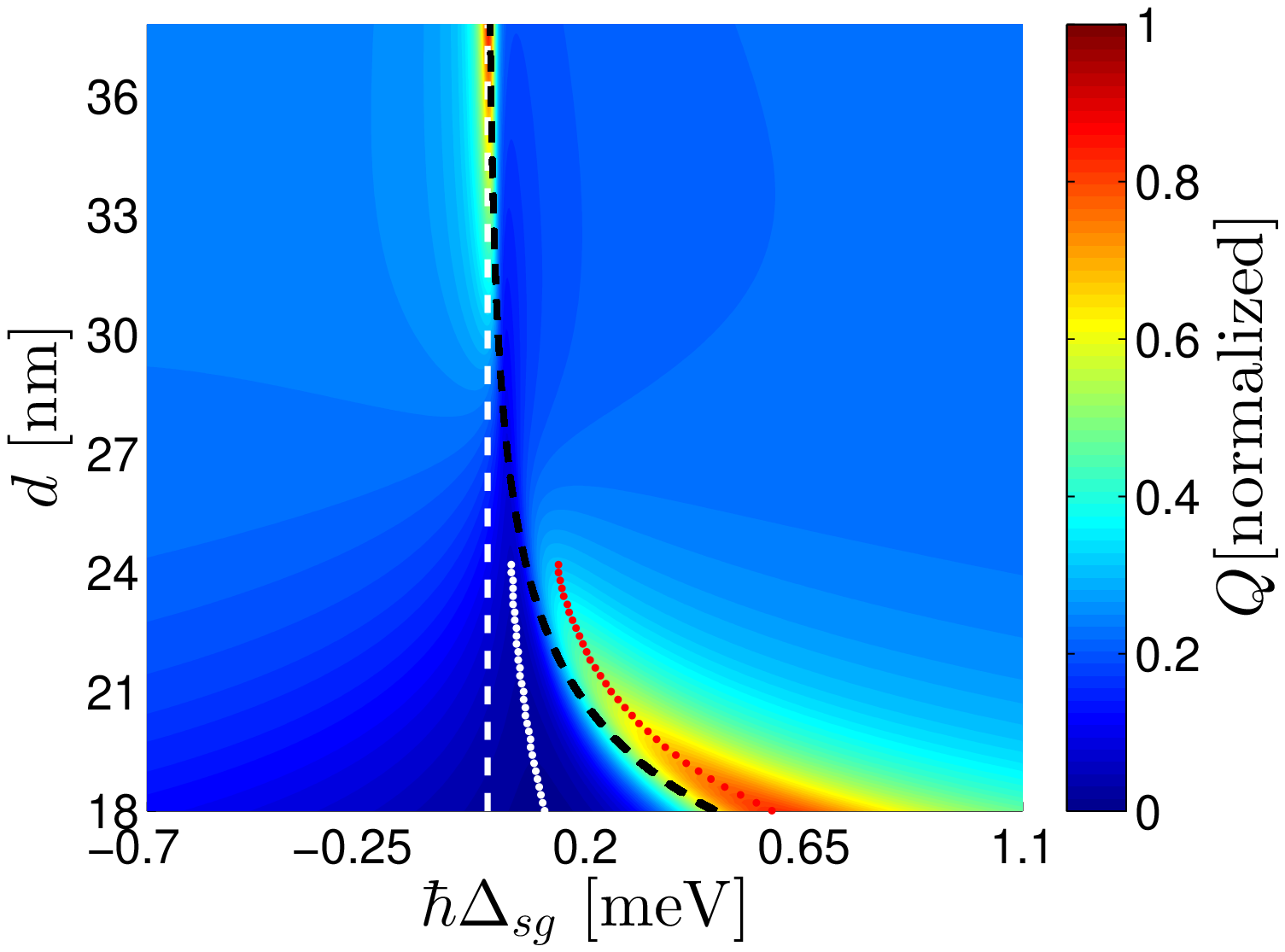}
\caption{Contour plot of the dimer-MNP hybrid absorption spectra $Q$ in the vicinity of the isolated dimer resonance $\omega_{sg}$ calculated for different center-to-center distances $d$. Decreasing $d$ corresponds to increasing the dimer-MNP coupling. The dashed black curve shows the $d$-dependence of the resonance frequency blue shift $\omega_{sg} + G_R$ \, ($G_R > 0$). The red and the white dotted lines highlight the position of the peak and the dip, respectively. The white dashed vertical line denotes the resonance frequency of the isolated dimer $\omega_{sg}$. The set of parameters are: $\hbar\omega_0=3.32~\mathrm{eV}$ (above the plasmon resonance), $\Gamma^{\prime} = 1.08 ~\gamma$ ($\hbar\Gamma^{\prime} = 10^{-3}~\mathrm{meV}$).}
\label{SpectLinearVard}
\end{figure}

The role of the ratio $|G|/\Gamma$ may also be investigated by varying $G$. Figure~\ref{SpectLinearVard} illustrates the crossover from the weak to the strong coupling regime upon changing the dimer-MNP distance $d$ from large to small values. In the calculations, we used the same set of parameters as in Fig~\ref{SpectLinearFano}(e) (we are thus in the high-energy plasmon tail), except that the pure dephasing rate was fixed at $\Gamma' = 1.08~\gamma$ ($\hbar \Gamma' = 10^{-3}~\mathrm{meV}$) and the dimer-MNP distance was varied from $d = 38 ~ \mathrm{to}~18~\mathrm{nm}$. As is seen from Fig.~\ref{SpectLinearVard}, at large dimer-MNP distances, the absorption spectrum of the dimer-MNP composite displays a narrow peak (bright red) which coincides with the resonance of an isolated dimer (shown by the vertical white dashed line), and its width is just $\Gamma^{\prime}$.

Upon decreasing $d$, the dimer peak weakens, because the dimer excitation starts to interact with the MNP and to share its energy with the plasmons. Around approximately $d = 25$ nm, the crossover to the strong coupling regime occurs and the system falls into the Fano regime. This is reflected in the absorption spectrum: a peak (red dotted line) and a dip (white dotted line) are clearly visible. Note also that upon decreasing $d$, the peak is blue-shifted with respect to the isolated dimer resonance and broadens. The black dashed curve highlights the $d$-dependence of the frequency shift.

Within a semiclassical picture (a quantum emitter coupled to a classical MNP), the shift and broadening have a natural explanation. Recall that the dimer resonance frequency $\omega_{sg}$ and the dephasing rate $\Gamma$ are renormalized due to the exciton-plasmon coupling: $\omega_{sg} \rightarrow \omega_{sg} + G_R Z_{sg}$ and $\Gamma \rightarrow \Gamma - G_I Z_{sg}$. Under the conditions considered here (weak excitation and blue-tail location of the dimer resonance with respect to the plasmon peak), both $G_R$ and $Z = -1$ are negative, while $G_I$ is always positive [see plots (b), (d), and (f) in Fig.~\ref{SpectLinearFano}]. This indeed predicts a blue-shifted renormalized frequency and an increase of the line width; moreover, both effects increase when $d$ decreases.

\subsection{Strong field regime}
\label{NonLinSpect}

In this section, we address the spectrum of the dimer-MNP hybrid in the limit of a strong electric field, when the two-photon transition from $|g \rangle \rightarrow |e \rangle$ comes into play. The hybrid's parameters were chosen as in Sec.~\ref{LinSpect}, with a fixed center-to-center distance $d = 18~\mathrm{nm}$ and restricting to the dimer's dephasing rate $\Gamma^{\prime} = 1.08~\gamma$ \, ($\hbar\Gamma^{\prime} = 10^{-3}~\mathrm{meV}$).
\begin{figure}[h!]
\centering
\includegraphics[width=0.80\linewidth]{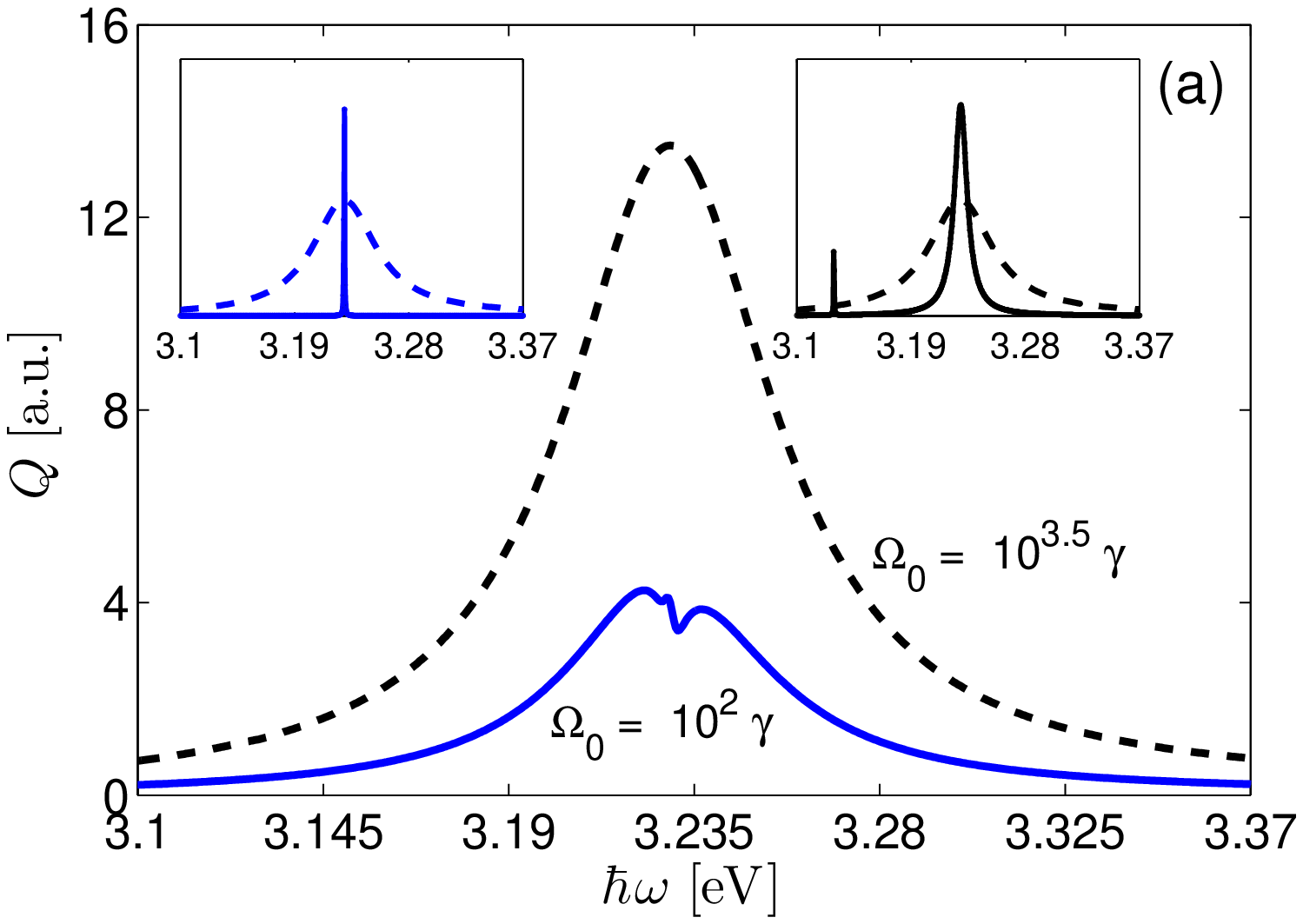}
\includegraphics[width=0.80\linewidth]{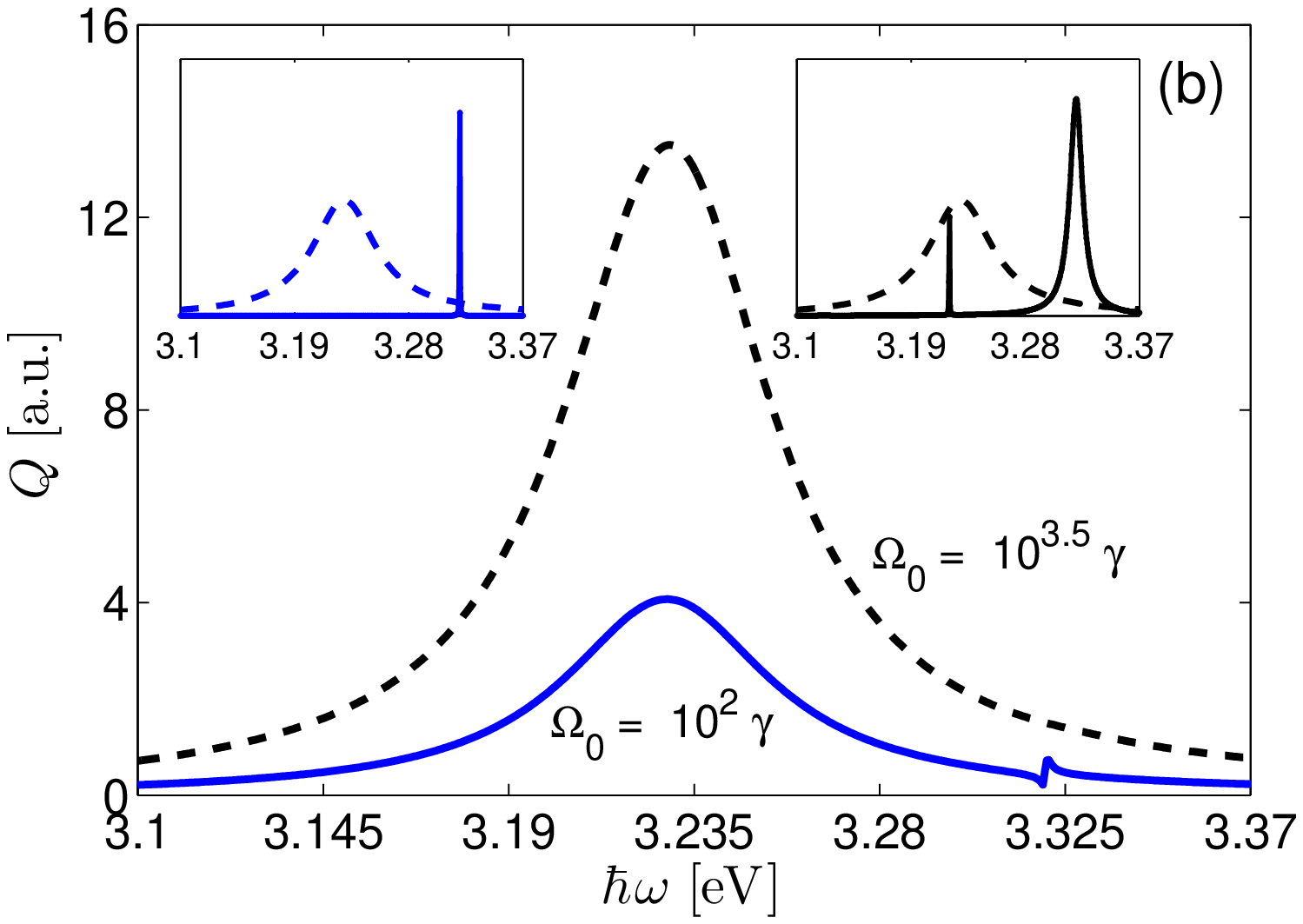}
\caption{Absorption spectra of the dimer-MNP hybrid in the strong field regime calculated for two values of the applied field strength: $\Omega_0 = 10^{2}\gamma$ (blue solid curve) and $\Omega_0 = 10^{3.5}\gamma$ (black dashed curve). Two dimer resonance frequencies are considered: (a)~$\hbar\omega_{sg} = 3.23~\mathrm{eV}$ (at the plasmon resonance) and (b)~$\hbar\omega_{sg} = 3.32~\mathrm{eV}$ (above the plasmon resonance). Inserts show spectra of the isolated dimer (solid curve) and of the MNP (dashed curve) for $\Omega_0 = 10^{2}\gamma$ (left insert) and $\Omega_0 = 10^{3.5}\gamma$ (right insert).}
\label{SpectHigh}
\end{figure}

Figure~\ref{SpectHigh} displays the absorption spectrum of the hybrid calculated for two values of the external field magnitude: $\Omega_0 = 10^{2}\gamma$ (blue solid curve) and $\Omega_0 = 10^{3.5}\gamma$ (black dashed curve). For $\Omega_0 = 10^{2}\gamma$, the system still exhibits a dip in the absorption when the frequency $\omega_{sg}$ of the isolated dimer's $|g \rangle \rightarrow |s \rangle$ transition coincides with the MNP plasmon resonance $\omega_{sp}$ [Fig.~\ref{SpectHigh} (a)]. However, the dip is much less pronounced than in the weak field limit (see Fig.~\ref{SpectLinearFano}), and does not drop to zero absorption anymore. This occurs due to saturation of the dimer's $|g \rangle \rightarrow |s \rangle$ transition, which also reduces the dimer-MNP interaction. The dip is also visible for $\omega_{sg} > \omega_{sp}$ [see Fig.~\ref{SpectHigh} (b)].

Upon increasing the applied field magnitude to $\Omega_0 = 10^{3.5}\gamma$, the dip in the hybrid's absorption is completely smeared, independently of the excitation frequency. This results from an even stronger saturation of the  dimer's $|g \rangle \rightarrow |s \rangle$ transition, which also experiences power broadening. At these strong fields, an additional peak in the isolated dimer spectra appears at the frequency of the two-photon transition $\hbar\omega_0$, which equals, respectively, 3.13 eV [Fig.\ref{SpectHigh} (a), right insert] and 3.22 {eV} [Fig.\ref{SpectHigh}(b), right insert]. This additional peak does not produce any discernible feature in the hybrid's absorption spectrum, meaning that the direct two-photon transition does not interact strongly with the MNP plasmon resonance. This results from the fact that this transition is only allowed in second-order perturbation theory, leading to a small effective transition dipole.

%%%
%%%

\section{Summary}
\label{summary}

In this work, we theoretically studied the optical response of a nanohybrid comprising a symmetric quantum dimer in close proximity to a MNP. The symmetric dimer consists of two identical quantum emitters (two-level molecules) and acts as an optical three-level system, with a ground state $|g \rangle$ (both emitters in the ground state), an optically active one-exciton state $|s \rangle$ (the symmetric linear combination of states in which one emitter is excited and the other one is in the ground state), and a doubly excited state $|e \rangle$ (both emitters excited). The antisymmetric one-exciton state is optically dark and may be ignored. The doubly excited state is optically active, however, it only can be excited via a two photon process, either through the one-exciton state as intermediate or by the direct absorption of two photons. We derived equations for the dimer's $3 \times 3$ density matrix describing its optical dynamics in the presence of a metal nanoparticle and analyzed these equations to study the optical properties of the hybrid.

We found that in a certain range of system parameters, the hybrid may exhibit bistability, but that this effect only occurs for the one-exciton transition, not for the two-exciton transition. The mechanism of this effect is identical to that of a single quantum emitter close to an MNP.~\cite{NugrohoJCP2013} The two-exciton transition exhibits excitation enhancement which causes saturation of the hybrid system at field intensities significantly lower than for an isolated dimer. The later indicates enhancement of the two photon absorption as has been observed experimentally.~\cite{SivapalanLang2012} No multi-stability occurs.

Another fascinating property of the dimer-MNP hybrid is its Fano-like absorption spectrum, with a shape that strongly depends on the spectral location of the dimer's one-exciton absorption peak with respect to the MNP's plasmon resonance and the ratio of coupling parameter and the dephasing rate. In particular, its basic shape depends on whether the exciton resonance lies at the low or the high frequency side of the plasmon resonance, or just at its maximum. In all cases the spectrum shows almost zero absorption at some frequency, making the hybrid transparent at this point. This effect is especially pronounced at the maximum of the MNP plasmon resonance. Here, the hybrid absorption shows a narrow dip to almost exactly zero.

Upon increasing the applied field intensity, the dip becomes less pronounced and disappears when the hybrid gets saturated. Similar effects occur if the one-exciton energy lies in one of the wings of the plasmon peak. In addition, the shape of the lineshape is sensitive to the coupling between the dimer's exciton transition and the plasmons. All this allows one to tailor the Fano-like shape of the hybrid's absorption spectrum by controlling the spectral and spatial separation of both constituents and by changing the input power.

We finally note that in this paper, we considered the limit where the magnitude of the coupling parameter $G$ is smaller than the spectral detuning between the  $|g \rangle \rightarrow |s \rangle$ and the $|s \rangle \rightarrow |e \rangle$ transitions of the quantum dimer. Preliminary calculations show that in the opposite limit, the hybrid shows a rich nonlinear dynamics, including auto-oscillations and chaos. This will be the subject of a forthcoming article.

\begin{acknowledgments}
\label{ack}
We are grateful to Prof. Vladimir Chernyak for helpful discussions. This work has been supported by NanoNextNL, a Micro- and Nano-technology consortium of the Government of the Netherlands and 130 partners.
\end{acknowledgments}

\begin{appendix}
\section{Size quantization of plasmons in Ag MNPs}
\label{ApendixA}

Here, we address the question of size quantization of plasmons confined in a small volume. The importance of size effects for an MNP can be estimated from comparing the energy level spacing in the vicinity of the Fermi energy $\varepsilon_F$ with the level width $\Gamma_p$ and thermal energy $k_B T$ ($k_B$ is the Boltzman constant). The spacing is given by~\citep{Kittel2005}
\begin{equation}
\Delta\varepsilon = (2/3) \varepsilon_F/N~,
\end{equation}
where $N$ is the number of electrons in a nanoparticle.  For an Ag nanoparticle, $\varepsilon_F = 5.48$ eV and the electron density $N/V_{MNP} = 5.85\times 10^{28}~ \mathrm{m}^{-3}$, where $V_{MNP}$ is the Ag nanoparticle volume.
Then, for a spherical Ag nanoparticle of size $r = 11$ nm (our case), we get $N = 3.26\times 10^5$ and, respectively, for the energy spacing $\Delta\varepsilon \approx 0.01$ meV. The thermal energy  $k_B T \approx 25$ meV (at room temperature), whereas $\Gamma_p \approx 0.1~\mathrm{fs}^{-1} \approx 65~\mathrm{meV}$.~\cite{maier2007}  Thus, both $k_BT$ and $\Gamma_p$ are much larger than $\Delta\varepsilon$, making the size quantization of confined plasmons irrelevant in our case. In a recently published paper,~\cite{SchollNat2012} it has been shown experimentally that size effects in Ag nanoparticles only play a role for radii smaller than to approximately 5 nm.

\section{Susceptibilities $\chi_{DIM}$ (dimer) and $\chi_{MNP}$ (MNP) at $\omega_{sg} \rightarrow \omega_{sp}$ in the weak-field regime}
\label{ApendixB}

In the weak field regime ($\Omega^0_{21} \ll \gamma_{21}$), the $|s \rangle \rightarrow |e \rangle$ transition is not involved in the dimer-MNP hybrid absorption. Moreover, the ground state depletion in this limit is negligible, meaning that $\rho_{gg} \approx 1$, $\rho_{ss} \approx 0$. Thus, the only relevant equation in this case is Eq.~(\ref{dRsg2}) for the coherence $R_{sg}$, which now takes the form
\begin{equation}
\dot{R}_{sg}  = -\left[ \Gamma + G_I - i(\Delta_{sg} + G_R) \right]R_{sg} - i\widetilde{\Omega}_0 \ .
\label{RsgLinear}
\end{equation}
We are interested in the steady-state solution to this equation, which reads
\begin{equation}
R_{sg}  = \frac{i}{ \Gamma + G_I - i(\Delta_{sg} + G_R) } \left(1 + \frac{\alpha}{2\pi d^3} \right) \Omega_0 \ ,
\label{RsgLinearSteadyState}
\end{equation}
where we used $\widetilde{\Omega}_0 = [ 1 + \alpha/(2\pi d^3)] \Omega_0$.
Having obtained the analytical solution for $R_{sg}$, one can write down the expressions for the dimer and MNP susceptibilities $\chi_\mathrm{D}$ and $\chi_\mathrm{MNP}$, the imaginary parts of which determine the absorption spectra.

First, we calculate $\chi_\mathrm{MNP}$. From Eqs.~(\ref{chiMNP}) and (\ref{RsgLinearSteadyState}), it is obtained as
\begin{equation}
\chi_\mathrm{MNP}  = \frac{1}{2V_{MNP}} \epsilon_b \alpha \left[  1 + \frac{iD^2}{2\pi \hbar \epsilon_0 \epsilon_b d^3} \,\,
\frac{1}{ \Gamma + G_I - i(\Delta_{sg} + G_R)}  \left(1 + \frac{\alpha}{2\pi d^3} \right) \right]  \ .
\label{ChiMNP}
\end{equation}
The general expression for $\mathrm{Im[\chi_{MNP}]}$ is bulky and hard to analyse analytically. However, at the maximum of the MNP plasmon peak, $\omega = \omega_{sp}$, and assuming perfect resonance between the dimer and the MNP plasmon, $\omega_{sg} = \omega_{sp}$, one can derive a simple formula for $\mathrm{Im[\alpha]}$. Realizing that under these conditions both $\mathrm{Re[\alpha]}$ and $G_R$ are equal to zero [see Fig.~\ref{SpectLinearFano}(d)], we obtain
\begin{equation}
\mathrm{Im}[\chi_{MNP}]  = \frac{\epsilon_b\alpha_I}{2V_{MNP}}\left[ 1 - \frac{D^2\alpha_I}{4\pi^2\hbar\epsilon_0\epsilon_b d^6(\Gamma + G_I)} \right] \ ,
\label{ChiMNPatSPR1}
\end{equation}
where $\alpha_I = \mathrm{Im[\alpha]}$. Because the factor that multiplies $1/(\Gamma+G_I)$ in the second term within square brackets is nothing else than $G_I$, one finally gets
\begin{equation}
\mathrm{Im}[\chi_{MNP}]  = \frac{\epsilon_b\alpha_I}{2V_{MNP}}\,\frac{\Gamma}{\Gamma + G_I} \ .
\label{ChiMNPatSPR2}
\end{equation}

Within the same approximations as above, one obtains for the imaginary part of the dimer susceptibility
\begin{equation}
\mathrm{Im}[\chi_\mathrm{DIM}] = \frac{D R_{sg}}
                      {2V_{DIM} \varepsilon_0  E_\mathrm{0}} = \frac{D^2}
                      {4V_{DIM} \hbar \varepsilon_0}\, \frac{1}{\Gamma + G_I} \ .
\label{ChiDatSPR2}
\end{equation}
\end{appendix}

\bibliography{TheBibFile}
\bibliographystyle{apsrev4-1}

\end{document}